\documentclass[twocolumn]{aastex63} 
\usepackage{graphicx}

\newcommand{\nsample}{13 }

\begin{document}

\title{Visual Orbits \& Alignments of Planet Hosting Binary Systems}

\author[0000-0002-9903-9911]{Kathryn~V.~Lester}
\affiliation{NASA Ames Research Center, Moffett Field, CA 94035, USA}

\author[0000-0002-2532-2853]{Steve~B.~Howell}
\affiliation{NASA Ames Research Center, Moffett Field, CA 94035, USA}
 
\author[0000-0001-7233-7508]{Rachel~A.~Matson}
\affiliation{U.S. Naval Observatory, Washington, D.C. 20392, USA}

\author[0000-0001-9800-6248]{Elise~Furlan}
\affiliation{NASA Exoplanet Science Institute, Caltech/IPAC, Pasadena, CA 91125, USA}

\author[0000-0003-2519-6161]{Crystal~L.~Gnilka}
\affiliation{NASA Ames Research Center, Moffett Field, CA 94035, USA}

\author[0000-0001-7746-5795]{Colin~Littlefield}
\affiliation{NASA Ames Research Center, Moffett Field, CA 94035, USA}
\affiliation{Bay Area Environmental Research Institute, Moffett Field, CA 94035, USA}

\author[0000-0002-5741-3047]{David~R.~Ciardi}
\affiliation{NASA Exoplanet Science Institute, Caltech/IPAC, Pasadena, CA 91125, USA}

\author[0000-0002-0885-7215]{Mark~E.~Everett}
\affiliation{NSF’s National Optical-Infrared Astronomy Research Laboratory, Tucson, AZ 85719, USA}

\author[0000-0001-9309-0102]{Sergio~B.~Fajardo-Acosta}
\affiliation{Caltech/IPAC, Pasadena, CA 91125}

\author[0000-0002-2361-5812]{Catherine~A.~Clark}
\affiliation{NASA Exoplanet Science Institute, Caltech/IPAC, Pasadena, CA 91125, USA}
\affiliation{Jet Propulsion Laboratory, California Institute of Technology, Pasadena, CA 91109 USA}

\accepted{\today \ in The Astronomical Journal}

\begin{abstract}
 
Roughly half of Solar-type planet hosts have stellar companions, so understanding how these binary companions affect the formation and evolution of planets is an important component to understanding planetary systems overall. Measuring the dynamical properties of planet host binaries enables a valuable test of planet formation in multi-star systems and requires knowledge of the binary orbital parameters. Using high resolution imaging, we have measured the relative astrometry and visual orbits of \nsample binary systems where one of the stars is known to host a transiting exoplanet. Our results indicate that the mutual inclination between the orbits of the binary hosts and the transiting planets are well aligned. Our results for close binary systems ($a<100$~AU) complement past work for wide planet host binaries from Gaia.

\end{abstract}

\section{Introduction} 

Multi-star systems make up about 50\% of Solar-type stars \citep{raghavan10} and 25\% of M-type stars \citep{winters19} in the Solar neighborhood. Recent work has shown that the fraction of planet hosting stars with stellar companions is similar to that of field binaries \citep{horch14, matson18, clark22}, so understanding how stellar companions affect the formation and evolution of exoplanets is an important component to understanding planetary systems overall. Observational radial velocity surveys and transit detections for exoplanets both have biases against the study of binary star systems and their planets; radial velocity studies often avoid known binaries due to contamination from the companion's spectral lines \citep[e.g.,][]{chontos22}, while transit studies often miss terrestrial-size planets when flux dilution from stellar companions causes a transit to become shallower than the detectability of the survey \citep{lester21}. Therefore, our knowledge of planetary architectures, characteristics, and occurrence rates is biased toward single-star systems, despite the fact that a significant fraction of binary systems are likely to host exoplanets.

Theoretical studies show that a close stellar companion can impact planets through the truncation or misalignment of the protoplanetary disk \citep{artymowicz94, kraus12, martin14}, the formation and migration of gas giant planets \citep{dawson18, fontanive19}, and the scattering of planets in unstable triple star systems \citep{thebault15}. For example, recent simulations of protoplanetary disks around the primary star in wide binary systems (with separations $a = 100 - 400$ AU) often result in the disk fragmentation needed to form giant planets \citep{cadman22}. Modeling also predicts that the shape and size of the companion's orbit can play a significant role in planet formation, such that close, eccentric, or highly inclined companions could hinder planet formation  \citep{holman99, quintana02, jc15, cadman22}. 

Over the past decade, observational evidence has accumulated to indicate that planet formation is suppressed in close ($a<100$ AU) binary systems. First, high resolution imaging surveys of known transiting planet host stars from Kepler, K2, and TESS have found a dearth of close stellar companions \citep{bergfors13, wang14a, kraus16, fontanive19, moe21, lester21, fontanive21}. Next, when searching for planets in binary systems, the frequency of giant planets in close binaries was found to be significantly less than the frequency in wide ($a>$100 AU) binaries \citep{wang14b, hirsch21}. \citet{su21} also found that multi-planet systems are more often found in wide binaries. Furthermore, observations of young binaries show that protoplanetary disks are smaller and less massive in binaries than around single stars \citep{zurlo21}, suggesting that stellar companions within about 300~AU often truncate the protoplanetary disks \citep{harris12}. However, the detection of planets in systems with close companions \citep[e.g.,][]{hatzes03, dupuy16, winters19b} demonstrates it is possible for planets to form in such systems, so it is currently unclear why some close binaries are able to host planets and which factors influence the survival of the planet. 

Little observational evidence exists to test how the other binary orbital parameters (such as inclination and eccentricity) affect planet formation, primarily due to the high angular resolution and long time baselines required to measure the binary orbits. Several recent papers \citep{dupuy22, behmard22, christian22} began probing the mutual inclination between transiting planets and stellar companions and found that the orbital planes of the host binaries are often well aligned with the planetary orbits. For example, \citet{christian22} studied wide binaries from Gaia that likely formed through turbulent fragmentation, which results in protoplanetary disks randomly aligned with the stellar companion \citep{offner10}. They concluded that subsequent gravitational interactions with a close companion could re-align the protoplanetary disk and produce the observed alignments. 

Long term observational monitoring of planet host binaries is necessary to determine how host multiplicity and binary orbital properties influence planet formation. For this purpose, we present the first results from our astrometric monitoring campaign of planet host binaries. In this paper, we explore the mutual orbital alignment of close binary systems ($a<100$ AU) known to host at least one transiting planet, in order to help characterize the architectures of binary sytems with planets and help place constraints on the formation and evolutionary models. We present orbital inclinations and preliminary visual orbits of \nsample binaries hosting circumstellar (S-type) planets to test if these systems also show planet-binary alignment. We describe our sample and observations in Sections~\ref{secsample} and \ref{observations}, our visual orbit analysis in Section~\ref{orbits}, planet-binary alignment results in Section~\ref{results}, and our conclusions in Section~\ref{conclusion}. 

\break


\section{Sample}\label{secsample}

\begin{deluxetable*}{lccccccccc}
\tablewidth{0pt}
\tabletypesize{\footnotesize}
\tablecaption{Sample of Planet Host Binaries \label{sample} }
\tablehead{
\colhead{Target} 
& \colhead{TIC ID} 
& \colhead{$T_{\rm eff~A}$ (K)} 
& \colhead{$T_{\rm eff~B}$ (K)} 
& \colhead{$q$} 
& \colhead{$M_{tot}$ ($M_\odot$)} 
& \colhead{$\pi$ (mas)} 
& \colhead{$\mu_{RA}$ (mas/yr)} 
& \colhead{$\mu_{DEC}$ (mas/yr)} 
& \colhead{Reference} 
}
\startdata
KOI 270        & 270779644 & 5650 & 5340 & 0.90 & 1.90 & $3.84\pm0.05$    & $ -9.40\pm0.01$ &  $-44.40\pm0.01$  & 1,3  \\   
KOI 307        & 138097531 & 6000 & 5800 & 0.95 & 2.16 & $1.27\pm0.11$    & $ -4.10\pm0.12$ &  $ -3.88\pm0.12$  & 1,2  \\   
KOI 1613       & 120576846 & 6080 & 5340 & 0.75 & 2.10 & $2.03\pm0.50$    & $-18.78\pm0.54$ &  $-20.46\pm0.58$  & 1,2,3  \\   
KOI 1961       & 158552426 & 5350 & 5140 & 0.89 & 1.70 & $2.47\pm0.03$    & $  1.13\pm0.03$ &  $-22.97\pm0.03$  & 3  \\   
KOI 2124       &  27457135 & 4060 & 3300 & 0.50 & 0.90 & $3.33\pm0.04$    & $-12.85\pm0.06$ &  $-18.33\pm0.06$  & 1,3  \\   
KOI 3234       & 164525743 & 6350 & 6000 & 0.83 & 2.44 & $1.55\pm0.06$    & $ -3.24\pm0.07$ &  $-10.63\pm0.08$  & 1  \\   
KOI 3456       & 137408775 & 5600 & 5500 & 0.98 & 1.92 & $2.05\pm0.04$    & $  6.32\pm0.04$ &  $  0.41\pm0.05$  & 1  \\   
KOI 4252       & 158489110 & 3930 & 4000 & 0.83 & 1.10 & $5.08\pm0.02$    & $  4.41\pm0.03$ &  $ 25.69\pm0.03$  & 3  \\   
KOI 5971       &  27778479 & 4620 & 4300 & 0.94 & 1.36 & $2.52\pm0.02$    & $  9.27\pm0.03$ &  $ 28.23\pm0.03$  & 1  \\   
TOI 271        & 259511357 & 6110 & 3800 & 0.47 & 1.68 & $10.01\pm0.13$   & $ 46.72\pm0.15$ &  $ 49.46\pm0.17$  & 4  \\   
TOI 1287       & 352764091 & 5890 & 4500 & 0.71 & 1.80 & $10.76\pm0.03$   & $ 33.87\pm0.02$ &  $-88.51\pm0.02$  & 4,5 \\   
EPIC 212303338 & 422290347 & 5100 & 4410 & 0.79 & 1.54 & $12.48\pm0.09$   & $ 31.54\pm0.10$ &  $ 10.40\pm0.06$  & 6  \\   
EPIC 220555384 & 406410648 & 4160 & 4330 & 0.93 & 1.37 & $6.85\pm0.51$    & $ 28.90\pm1.45$ &  $-24.37\pm1.22$  & 6  \\   
\enddata  
\tablerefs{
1. \citet{furlan17},
2. \citet{colton21}
3. \citet{dupuy22},
4. \citet{lester21},
5. \citet{howell21},
6. \citet{matson18}.
}
\end{deluxetable*}

\begin{deluxetable*}{lccccc}
\tablewidth{0pt}
\tabletypesize{\footnotesize}
\tablecaption{Estimated Planet Properties \label{planets} }
\tablehead{
\colhead{Planet} 
& \colhead{$P_{pl}$ (days)} 
& \colhead{$R_{pl}$ ($R_\oplus$)} 
& \colhead{$a_{pl}$ (AU)} 
& \colhead{Designation\tablenotemark{*}} 
& \colhead{Reference} 
}
\startdata
KOI 270.01        & 12.6    & 1.53  & 0.10   & CP (Kepler-449 b)   &  1,2   \\  
KOI 270.02        & 33.7    & 1.86  & 0.20   & CP (Kepler-449 c)   &  1,2   \\
KOI 307.01        & 19.7    & 1.78  & 0.14   & CP (Kepler-520 b)   &  1,3    \\  
KOI 307.02        &  5.2    & 1.18  & 0.06   & CP (Kepler-520 c)   &  1,3  \\  
KOI 1613.01       & 15.9    & 1.31  & 0.12   & CP (Kepler-907 b)   &  1,3    \\
KOI 1613.02       & 20.6    & 0.85  & 0.15   & Candidate           &  4    \\
KOI 1613.03       & 94.1    & 0.90  & 0.40   & Candidate           &  4    \\
KOI 1961.01       &  1.9    & 0.91  & 0.03   & CP (Kepler-1027 b)  &  1,3   \\
KOI 2124.01       & 42.3    & 1.45  & 0.20   & Candidate           &  4    \\
KOI 3234.01       &  2.4    & 0.83  & 0.04   & CP (Kepler-1443 b)  &  3,4    \\
KOI 3456.01       & 30.9    & 1.08  & 0.19   & CP (Kepler-1505 b)  &  3,4    \\
KOI 3456.02       & 486.1   & 1.18  & 1.20   & Candidate           &  5    \\      
KOI 4252.01       &  15.6   & 0.72  & 0.10   & CP (Kepler-1948 b)  &  4,6    \\
KOI 5971.01       & 493.3   & 1.08  & 1.00   & Candidate           &  5    \\
TOI 271.01        &   2.5   & 2.81  & 0.04   & Candidate           &  7    \\
TOI 1287.01       &   9.6   & 2.52  & 0.09   & Candidate           &  7   \\
EPIC 212303338.01 &   0.6   & 0.58  & 0.01   & Candidate           &  8,9    \\
EPIC 220555384.01 &   4.3   & 1.20  & 0.05   & Candidate           &  8,10    \\
\enddata  
\tablenotetext{*}{We note whether each planet is listed on the NASA Exoplanet Archive as a confirmed planet (CP) or planet candidate.}
\tablerefs{
1. \citet{batalha13}, 
2. \citet{ve15},
3. \citet{morton16},
4. \citet{q1-12table}, 
5. \citet{q1-16table}, 
6. \citet{valizadegan22},
7. \citet{toi},
8. \citet{epic},
9. \citet{kostov19},
10. \citet{kruse19}.
}
\end{deluxetable*}

We started building our sample from transiting planet host stars from Kepler, K2, and TESS for which close stellar companions were previously detected using speckle interferometry \citep{furlan17, matson18, ziegler20, howell21, lester21}. For each binary, we estimated the projected physical separation using the projected angular separation from the most recent speckle epoch and the Gaia DR3 parallax \citep{gaia, gaiaDR3}. We then kept only those with projected separations less than 100 AU, where stellar companions are most likely to impact planet formation. Transit false positive systems identified by follow-up photometry as listed on the Exoplanet Follow-up Observing Program (ExoFOP) website were also removed. Our full sample contains 40 binaries, for which we are conducting an on-going astrometric and spectroscopic monitoring campaign with the Gemini, WIYN, and Keck telescopes. With angular separations less than 1~arcsec, these systems are expected to be gravitationally bound \citep{everett15, hirsch17, matson18} but we confirm the bound nature of each system in Section~\ref{orbits}.

We present preliminary visual orbit solutions for \nsample of the exoplanet host binaries in our sample, for which orbital motion can already be seen. These systems are listed in Table~\ref{sample} with their TIC ID, primary star effective temperature ($T_{\rm eff~A}$) from the TIC catalog,  estimates of the secondary star effective temperature ($T_{\rm eff~B}$), binary mass ratio ($q$) and total system mass ($M_{tot}$) from the magnitude difference (see Section~\ref{orbits}), Gaia DR3 parallax ($\pi$) and proper motion ($\mu$), and companion detection reference. We list the planet properties in Table~\ref{planets}, including the planet name, period ($P_{pl}$), radius ($R_{pl}$, uncorrected for flux dilution) and semi-major axis ($a_{pl}$) from the KOI/EPIC/TOI catalogs, whether each planet is designated on the NASA Exoplanet Archive as a planet candidate or confirmed planet (CP), and literature reference. Most of the binaries discussed herein are Kepler targets due to the long observational baselines available, and therefore Table~\ref{sample} contains mainly Solar-type stars with Earth-sized planets. Four systems have multiple planets/planet candidates, so we list the estimated properties of each one.

\begin{figure*}
\centering
\includegraphics[width=\textwidth]{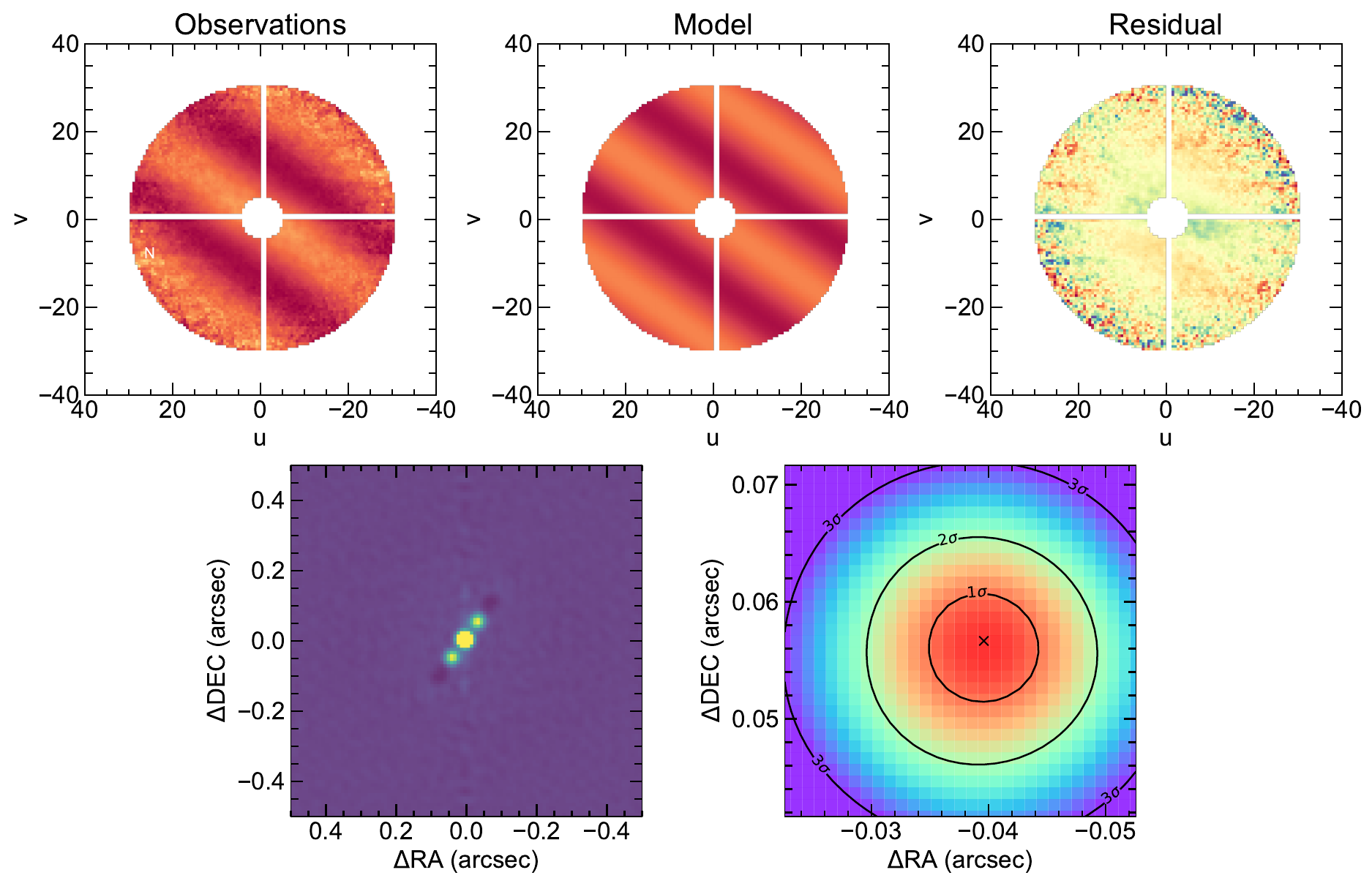}
\caption{Power spectra and relative astrometric solution for KOI~4252 on 2021 Oct 24. The top panel shows the observed binary power spectrum (left), the best-fit model (center), and the residuals (right). The bottom left plot shows the full reconsructed image from the speckle pipeline, which often has a reflected image of the companion. The bottom right plot shows a close-in view of the $\chi^2$ values around the best-fit solution, where the best-fit position of the companion is marked with an X. The 1-, 2-, and 3-$\sigma~\chi^2$ contour levels (corresponding to $\chi^2_{min}+1,~ \chi^2_{min}+4,~ \chi^2_{min}+9$) are shown in black. 
\label{gridfit}}
\end{figure*}

\begin{deluxetable*}{lccccccc}
\tablewidth{0pt}
\tabletypesize{\small}
\tablecaption{New Relative Astrometry from Speckle Interferometry\label{relpos}    }
\tablehead{
\colhead{Target} 
& \colhead{UT Date} 
& \colhead{MJD} 
& \colhead{$\rho$ (mas)} 
& \colhead{$\theta$ (deg)} 
& \colhead{$\Delta m$ (mag)} 
& \colhead{Filter} 
& \colhead{Telescope} 
}
\startdata
KOI 270        &  2022-09-16  &  59836.30  &  $186.0\pm 1.6$  &  $ 65.0\pm 0.6$  &  $0.6\pm0.5$  &  $i'$     &  WIYN   \\ 
KOI 1613       &  2022-09-19  &  59836.21  &  $191.5\pm 4.0$  &  $185.3\pm 0.6$  &  $1.2\pm0.5$  &  $i'$     &  WIYN   \\ 
KOI 1961       &  2021-06-24  &  59389.47  &  $ 46.8\pm 2.0$  &  $275.2\pm 2.4$  &  $0.2\pm0.4$  &  832~nm   &  Gemini \\ 
KOI 1961       &  2021-10-15  &  59502.25  &  $ 44.9\pm 2.5$  &  $276.7\pm 3.8$  &  $0.0\pm0.5$  &  832~nm   &  Gemini \\ 
KOI 1961       &  2022-05-10  &  59709.55  &  $ 47.0\pm 3.5$  &  $277.6\pm 3.7$  &  $0.4\pm0.3$  &  832~nm   &  Gemini \\ 
KOI 2124       &  2021-06-26  &  59391.53  &  $ 79.5\pm 3.3$  &  $ 53.3\pm 2.3$  &  $0.3\pm0.4$  &  832~nm   &  Gemini \\ 
KOI 2124       &  2021-10-19  &  59506.26  &  $ 78.1\pm 2.5$  &  $ 53.5\pm 1.8$  &  $0.2\pm0.3$  &  832~nm   &  Gemini \\ 
KOI 2124       &  2022-05-09  &  59708.61  &  $ 80.9\pm 5.2$  &  $ 53.2\pm 3.8$  &  $0.5\pm0.4$  &  832~nm   &  Gemini \\ 
KOI 3234       &  2022-09-14  &  59836.77  &  $ 70.5\pm 5.0$  &  $158.6\pm 3.0$  &  $0.9\pm0.5$  &  $z'$     &  WIYN   \\
KOI 3456       &  2022-09-12  &  59834.28  &  $ 50.8\pm 3.5$  &  $ 11.9\pm 3.9$  &  $0.0\pm1.3$  &  832~nm   & Gemini \\
KOI 4252       &  2021-06-25  &  59390.50  &  $ 67.7\pm 2.5$  &  $325.3\pm 2.1$  &  $0.6\pm0.2$  &  832~nm   &  Gemini \\ 
KOI 4252       &  2021-10-24  &  59511.21  &  $ 69.1\pm 4.2$  &  $325.0\pm 3.6$  &  $0.8\pm0.2$  &  832~nm   &  Gemini \\ 
KOI 4252       &  2022-05-09  &  59708.55  &  $ 70.3\pm 4.5$  &  $323.7\pm 3.7$  &  $0.8\pm0.2$  &  832~nm   &  Gemini \\ 
KOI 4252       &  2022-09-12  &  59834.28  &  $ 72.9\pm 3.5$  &  $323.2\pm 2.7$  &  $0.5\pm0.2$  &  832~nm   &  Gemini \\ 
KOI 5971       &  2021-06-28  &  59393.50  &  $ 29.9\pm 4.5$  &  $128.0\pm 8.6$  &  $1.0\pm0.9$  &  832~nm   &  Gemini \\ 
KOI 5971       &  2021-10-21  &  59508.25  &  $ 29.9\pm 3.3$  &  $128.0\pm 6.1$  &  $0.8\pm0.5$  &  832~nm   &  Gemini \\ 
KOI 5971       &  2022-05-11  &  59710.56  &  $ 26.9\pm 5.7$  &  $130.2\pm12.3$  &  $0.8\pm1.0$  &  832~nm   &  Gemini \\ 
TOI 271        &  2021-09-18  &  59840.77  &  $153.0\pm 5.0$  &  $226.8\pm 2.0$  &  $5.1\pm1.0$  &  832~nm   &  Gemini \\ 
TOI 1287       &  2021-06-24  &  59389.54  &  $131.5\pm 9.0$  &  $346.4\pm 3.9$  &  $3.2\pm0.5$  &  832~nm   &  Gemini \\ 
TOI 1287       &  2021-10-23  &  59510.24  &  $135.8\pm10.1$  &  $346.0\pm 4.6$  &  $3.3\pm0.6$  &  832~nm   &  Gemini \\ 
TOI 1287       &  2022-05-11  &  59710.58  &  $144.7\pm11.1$  &  $346.4\pm 4.7$  &  $3.3\pm0.7$  &  832~nm   &  Gemini \\ 
TOI 1287       &  2022-09-18  &  59840.51  &  $147.0\pm10.0$  &  $349.4\pm 5.0$  &  $2.7\pm1.0$  &  $z'$     &  WIYN   \\ 
EPIC 212303338 &  2023-01-28  &  59971.50  &  $124.0\pm 5.0$  &  $100.4\pm 2.0$  &  $1.8\pm0.5$  &  $z'$     &  WIYN   \\ 
EPIC 220555384 &  2021-10-16  &  59503.40  &  $210.9\pm 2.0$  &  $276.9\pm 0.5$  &  $0.7\pm0.1$  &  832~nm   &  Gemini \\ 
EPIC 220555384 &  2021-12-09  &  59557.25  &  $204.9\pm 5.0$  &  $277.1\pm 2.6$  &  $0.7\pm0.1$  &  832~nm   &  Gemini \\ 
EPIC 220555384 &  2022-09-14  &  59837.51  &  $211.5\pm 2.4$  &  $278.3\pm 0.5$  &  $0.7\pm0.3$  &  $i'$     &  WIYN   \\    
EPIC 220555384 &  2022-09-15  &  59837.51  &  $211.9\pm 2.5$  &  $276.5\pm 0.5$  &  $0.7\pm0.1$  &  $i'$     &  Gemini \\  
\enddata  
\end{deluxetable*}

\section{Observations} \label{observations}

We observed our binary sample using the `Alopeke and Zorro speckle cameras \citep{scott21} on the Gemini 8.1~m North and South telescopes from June 2021 to September 2022 and using the NN-EXPLORE Exoplanet and Stellar Speckle Imager (NESSI) speckle camera on the WIYN 3.5~m telescope \citep{scott18} from October 2022 to January 2023. At least three image sets were obtained for each target, where one set consists of 1000 60~ms (Gemini) or 40~ms (WIYN) exposures taken simultaneously in two filters. The 2021 data were taken using 562~nm and 832~nm narrow-band filters, while some of the 2022--2023 data were taken using the SDSS $r'$, $i'$, or $z'$ broad-band filters to increase the signal-to-noise ratio. Additional image sets were taken for fainter targets ($V>9$ mag), and a point source standard star was observed immediately before or after each target for calibration. We reduced the data using the pipeline developed by the speckle team \citep{howell11, horch11} to calculate the power spectrum of each target, divide the mean power spectrum of the target by that of the standard star, and fit the fringes for initial estimates of the binary parameters. For solutions with a 180~deg position angle ambiguity, we selected the solution consistent with other speckle or adaptive optics observations.

We then determined the final relative positions and uncertainties from the binary power spectra by performing a grid search in relative separation and position angle based on the gridfit code of \citet{schaefer16}. We first calibrated the $uv$-plane with the power spectra of known binary stars and the predicted relative positions from literature orbital solutions: HD~214850 and HIP~46454 \citep{muterspaugh10}, HIP~84949 \citep{muterspaugh06}, and HIP~4849 \citep{tokovinin15}. Once the $uv$-plane was calibrated for each observing run, we tested a range of separations in right ascension ($\Delta$RA) and declination ($\Delta$DEC) around the solution found by the speckle pipeline in steps of 1~mas. At each grid point, we created a model power spectrum for these binary parameters, fit for the magnitude difference of the binary, and calculated the $\chi^2$ goodness-of-fit statistic between the observed and model fringes. We then mapped out the $1\sigma$  $\chi^2$ contour, fit for the uncertainties in $\Delta$RA and $\Delta$DEC, and converted these values \& uncertainties to relative separation ($\rho$) and position angle ($\theta$, measured East of North). An example power spectrum, reconstructed image, and $\chi^2$ map are shown in Figure~\ref{gridfit}. Table~\ref{relpos} lists the UT date, Modified Julian Date (MJD), separation, position angle, magnitude difference, filter, and telescope for each observation.


\section{Visual Orbits} \label{orbits}
We combined our new relative astrometry with past measurements from Keck NIRC2 adaptive optics observations \citep{furlan17, dupuy22}, WIYN speckle observations \citep{matson18, colton21, howell21}, and Gemini speckle observations \citep{furlan17, lester21}.  If uncertanties were not listed in the literature, we adopted values of 5~mas and 2~deg for the relative separation and position angle, repsectively \citep{howell21}. 

\begin{figure}
\centering
\includegraphics[width=0.48\textwidth]{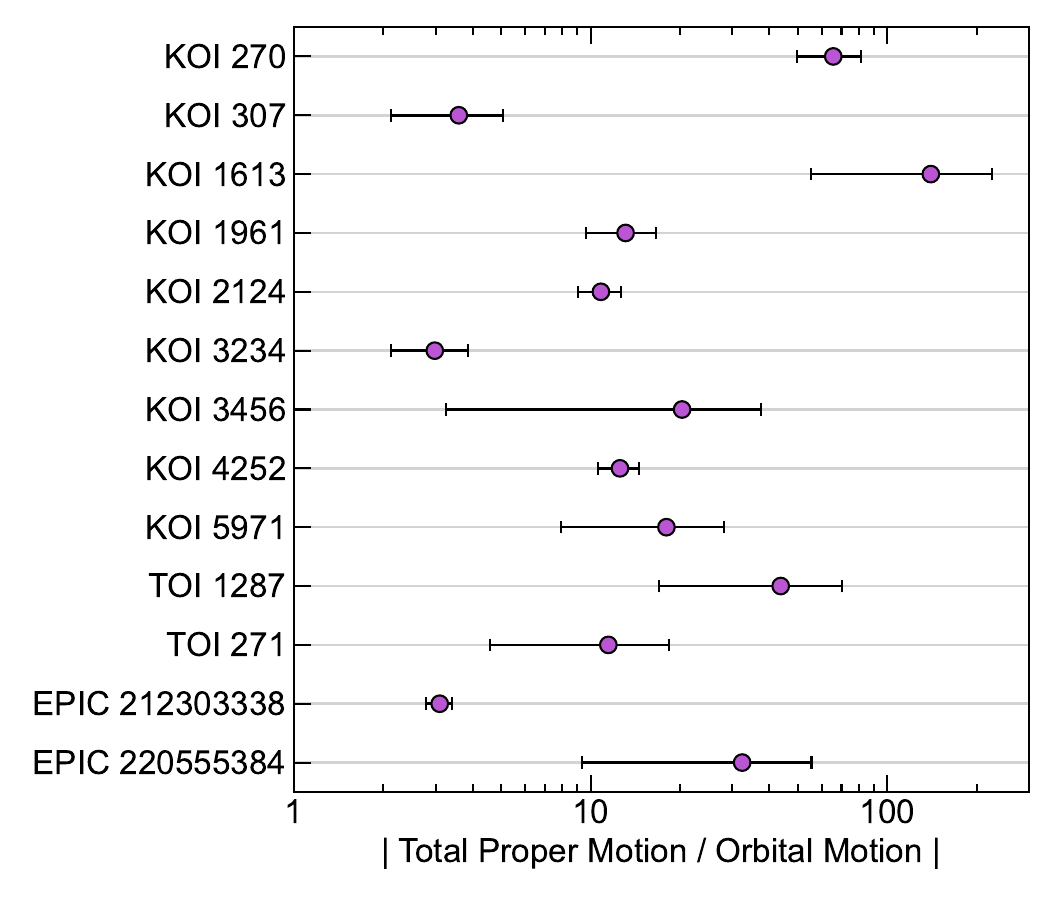}
\caption{Ratio of the total Gaia DR3 proper motion to the total observed motion for each binary in our sample. The proper motion is at least three times larger compared to the observed relative motion for all systems, so this motion is likely true orbital motion of a bound companion rather than motion of an unbound, line-of-sight companion. 
\label{pm}}
\end{figure}

We first used the compiled relative astrometry data to test the bound nature of each binary system and confirm that the observed on-sky motion is actually orbital motion. Because our binary stars are unresolved by Gaia, we could not do a typical common proper motion analysis \citep[e.g.,][]{colton21}. Instead, we compared the proper motion ($\mu$) of the primary star from Gaia DR3 (listed in Table~\ref{sample}) to the observed relative motion of the secondary star. If the binary companion is unbound, i.e. a background line-of-sight companion, then the companion's observed motion would be equal in magnitude to the proper motion of the primary star. Figure~\ref{pm} shows the ratio of the total proper motion to the mean angular speed of the companion, which was calculated in RA and DEC separately from the first to last observations then added in quadrature. We found that the proper motion was $3-140$ times larger compared to the observed motion for all our binaries. Therefore, the observed motion is likely true orbital motion and can be fit with a Keplerian orbit. We also show the direction of the primary star's proper motion in the orbit plots in the Appendix to compare with the orbital motion. 

\centerwidetable
\begin{deluxetable*}{lrrrrrrr}
\tabletypesize{\small}
\tablewidth{0pt}
\tablecaption{Visual Orbit Solutions \label{orbpar} }
\tablehead{
\colhead{Target} 
& \colhead{$P$ (yr)} 
& \colhead{$\tau$} 
& \colhead{$i$ (deg)} 
& \colhead{$e$ } 
& \colhead{$\omega_B$ (deg)} 
& \colhead{$\Omega_B$ (deg)} 
& \colhead{$a$ (AU)} 
}
\startdata
KOI 270        & $290.0^{+251.0}_{-200.0}$ & $ 0.65^{+0.10}_{-0.10}$ & $ 86.8^{+ 1.3}_{- 1.9}$ & $ 0.10^{+0.18}_{-0.10}$ & $323.5^{+ 8.7}_{- 7.2}$ & $248.8^{+ 3.5}_{- 3.5}$ & $ 55.0^{+11.1}_{-16.0}$ \\
KOI 307        & $290.0^{+212.8}_{- 75.5}$ & $ 0.03^{+0.08}_{-0.08}$ & $129.0^{+24.5}_{-23.7}$ & $ 0.67^{+0.25}_{-0.25}$ & $165.0^{+49.3}_{-54.4}$ & $267.5^{+25.0}_{-25.0}$ & $ 61.0^{+20.5}_{-14.9}$ \\
KOI 1613       & $675.0^{+557.4}_{-218.3}$ & $ 0.15^{+0.14}_{-0.14}$ & $ 86.5^{+ 3.9}_{- 4.5}$ & $ 0.23^{+0.19}_{-0.19}$ & $152.5^{+47.6}_{-53.4}$ & $182.5^{+ 3.3}_{- 3.3}$ & $102.5^{+38.6}_{-27.0}$ \\
KOI 1961       & $ 27.5^{+ 73.9}_{-  9.5}$ & $ 0.55^{+0.08}_{-0.08}$ & $ 64.5^{+ 6.6}_{-11.8}$ & $ 0.79^{+0.21}_{-0.22}$ & $325.0^{+27.7}_{-21.9}$ & $101.0^{+22.1}_{-22.1}$ & $ 11.5^{+ 3.1}_{- 2.3}$ \\
KOI 2124       & $150.0^{+243.4}_{-163.5}$ & $ 0.77^{+0.13}_{-0.13}$ & $ 89.9^{+ 1.2}_{- 1.3}$ & $ 0.04^{+0.32}_{-0.04}$ & $341.0^{+ 8.3}_{-11.9}$ & $233.8^{+ 1.8}_{- 1.8}$ & $ 27.0^{+12.0}_{- 9.3}$ \\
KOI 3234       & $175.0^{+362.5}_{- 54.7}$ & $ 0.97^{+0.29}_{-0.29}$ & $ 33.0^{+23.0}_{-31.7}$ & $ 0.01^{+0.53}_{-0.01}$ & $135.0^{+51.1}_{-59.9}$ & $196.0^{+42.9}_{-42.9}$ & $ 43.0^{+16.1}_{- 9.0}$ \\
KOI 3456       & $ 37.5^{+ 71.7}_{- 10.0}$ & $ 0.55^{+0.15}_{-0.15}$ & $ 97.0^{+10.3}_{-10.1}$ & $ 0.97^{+0.03}_{-0.63}$ & $312.5^{+21.7}_{-28.6}$ & $192.5^{+ 4.2}_{- 4.2}$ & $ 15.0^{+ 4.7}_{- 4.3}$ \\
KOI 4252       & $ 70.0^{+ 75.0}_{- 40.0}$ & $ 0.73^{+0.16}_{-0.16}$ & $ 99.5^{+ 5.5}_{- 3.5}$ & $ 0.33^{+0.10}_{-0.10}$ & $307.5^{+12.2}_{- 9.6}$ & $117.0^{+ 4.1}_{- 4.1}$ & $ 19.0^{+12.5}_{- 6.3}$ \\
KOI 5971       & $ 50.0^{+ 46.8}_{- 40.0}$ & $ 0.29^{+0.12}_{-0.12}$ & $ 93.0^{+ 6.2}_{- 6.2}$ & $ 0.39^{+0.28}_{-0.28}$ & $ 37.5^{+25.1}_{-32.3}$ & $313.0^{+ 4.8}_{- 4.8}$ & $ 13.0^{+ 4.5}_{- 4.7}$ \\
TOI 271        & $ 22.5^{+ 47.2}_{- 12.3}$ & $ 0.73^{+0.11}_{-0.11}$ & $ 98.5^{+11.0}_{- 6.2}$ & $ 0.95^{+0.05}_{-0.08}$ & $327.5^{+95.6}_{-24.9}$ & $ 49.8^{+ 4.2}_{- 4.2}$ & $ 11.0^{+ 4.3}_{- 3.4}$ \\
TOI 1287       & $ 27.5^{+ 39.6}_{- 13.3}$ & $ 0.75^{+0.11}_{-0.11}$ & $ 86.8^{+ 3.1}_{- 4.3}$ & $ 0.29^{+0.49}_{-0.29}$ & $341.0^{+36.8}_{-20.3}$ & $169.0^{+ 4.3}_{- 4.3}$ & $ 11.5^{+ 3.9}_{- 2.7}$ \\
EPIC 212303338 & $ 57.5^{+184.5}_{- 25.8}$ & $ 0.63^{+0.27}_{-0.27}$ & $103.5^{+ 8.1}_{- 4.0}$ & $ 0.47^{+0.14}_{-0.14}$ & $102.5^{+34.0}_{-32.6}$ & $ 44.5^{+ 5.8}_{- 5.8}$ & $ 17.5^{+14.3}_{- 5.1}$ \\
EPIC 220555384 & $ 67.5^{+231.0}_{- 16.2}$ & $ 0.59^{+0.12}_{-0.12}$ & $ 77.0^{+ 6.6}_{-16.1}$ & $ 0.91^{+0.09}_{-0.33}$ & $293.0^{+14.3}_{-22.2}$ & $ 99.0^{+ 3.4}_{- 3.4}$ & $ 18.5^{+11.3}_{- 3.2}$ \\
\enddata  
\end{deluxetable*} 

\begin{figure*}
\centering
\includegraphics[width=0.95\textwidth]{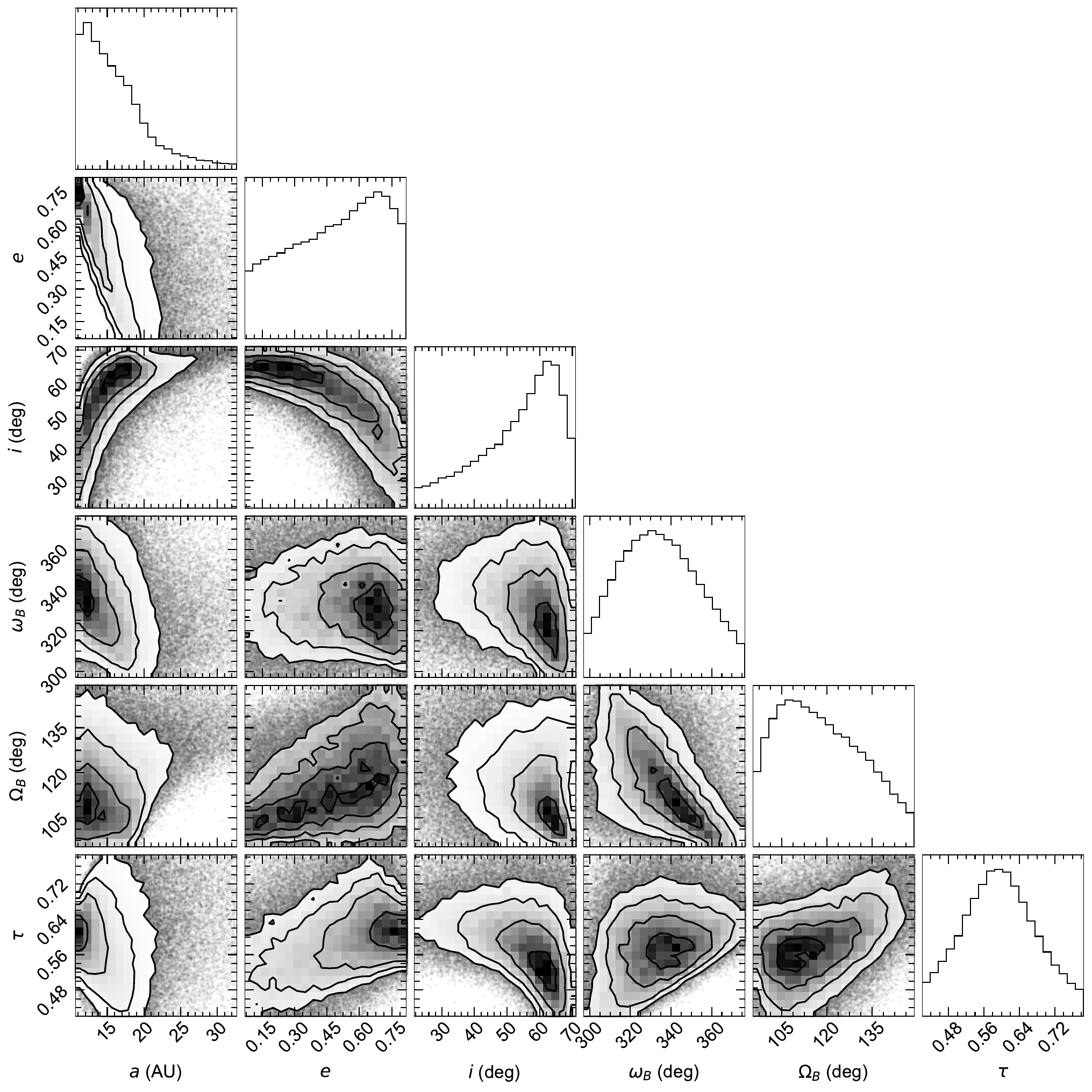}
\caption{Example corner plot of the orbital solution for KOI~1961. The diagonal frames show posterior histograms for each orbital parameter, and the off-diagonal frames show the covariance between different pairs of parameters.
\label{corner}}
\vspace{12pt}
\end{figure*}

\begin{figure}
\centering
\includegraphics[width=0.48\textwidth]{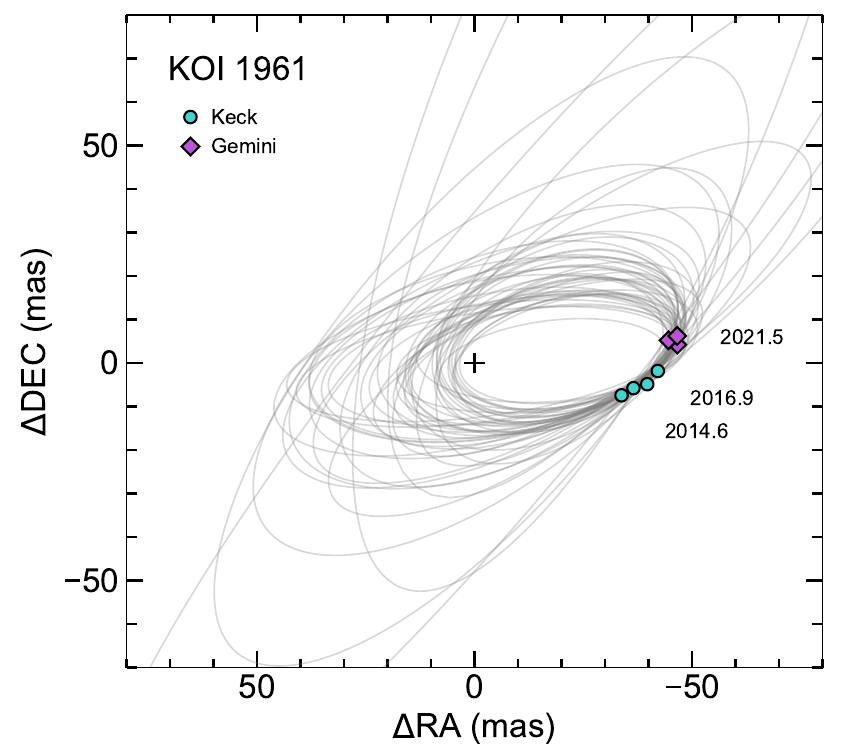}
\caption{Example visual orbit for planet hosting binary KOI~1961. The primary star is positioned at the origin (black cross), and the relative positions of the secondary component are marked with colored points. A random subset of the orbital solutions from \texttt{orbitize!} are shown in grey.
\label{vb}}
\end{figure}

We fit for the visual orbits using the \texttt{orbitize!} package \citep{orbitize} and Orbits For The Impatient \citep{ofti} module, which was built specifically for long period systems. For each binary, we estimated the primary star's mass from the effective temperature and the Modern Mean Dwarf Stellar Color and Effective Temperature Sequence \citep{pecaut13}, then used the speckle magnitude difference to estimate the secondary's mass \citep[see][]{matson18}. We used the resulting total mass (with uncertainties of 30\%) and the Gaia DR3 parallaxes \& uncertainties \citep{gaia} as priors, which are listed in Table~\ref{sample}.  The free parameters were then the semi-major axis ($a$), inclination ($i$), eccentricity ($e$), argument of periastron of the companion ($\omega_B$)
\footnote{Note that our visual orbit solutions use argument of periastron of the secondary star ($\omega_B$), while spectroscopic orbit solutions typically use the argument of periastron of the primary star ($\omega_A = \omega_B + 180^\circ$).}, 
longitude of the ascending node of the companion ($\Omega_B$), and epoch of periastron. \texttt{Orbitize!} uses a parameter $\tau$ to represent the epoch of periastron as a fraction of the orbital period past the reference epoch MJD 58849.  We ran \texttt{orbitize!} until $10^5$ orbits were accepted, created histograms for each orbital parameter, and fit asymmetrical Gaussians to each distribution to find the best-fit values and uncertainties. Table~\ref{orbpar} lists the orbital solutions for each binary. An example corner plot and visual orbit are shown in Figures~\ref{corner} and \ref{vb}, respectively, while the visual orbits for all systems are shown in Figures~\ref{vb_koi270}--\ref{vb_epic384} in the Appendix. Next, we used the total system mass and the semi-major axis to estimate the orbital period ($P$) for each system. Our observations cover roughly 1--25\% of the orbits so the orbital periods are not yet well constrained, but orbital coverage of a few percent is sufficient to reliably measure the orbital inclination \citep{dupuy22}.

As a consistency check, we also fit for the visual orbits using a custom code. We created $10^6$ sets of random orbital parameters, calculated the predicted binary positions, and determined the $\chi^2$ value of each solution. Orbital parameters for each iteration were drawn from uniform distributions. We then found parameters with the lowest reduced $\chi^2$ value, fit a parabola to the bottom of the $\chi^2$ distribution, and found the $1\sigma$ uncertainties where $\chi^2 \le \chi^2_{min} + 1$. The inclinations from \texttt{orbitize!} are consistent with those found by our fitting method to within the uncertainties. However, our code could not converge on a full orbital solution as well as \texttt{orbitize!} due to the orbital period as a free parameter, so we used the \texttt{orbitize!} solutions in the rest of this paper.

\section{Results \label{results}} 

\subsection{Planet-Binary Orbital Alignment}
\begin{figure}
\centering
\includegraphics[width=0.48\textwidth]{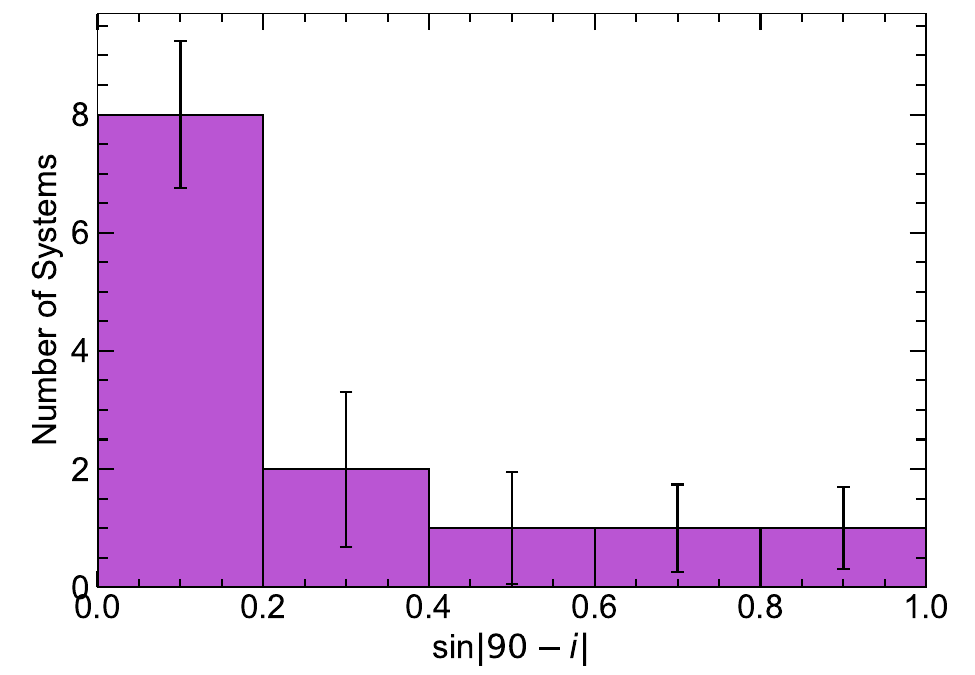}
\caption{Alignment between the planetary and binary orbital inclinations ($\sin|90-i|$). A random inclination distribution would result in uniform histograms, but instead our binary host star orbits are often well aligned with the planets' orbits. This is consistent with numerical simulations of planet formation, which found it more difficult to form planets in protoplanetary disks with a highly misaligned stellar companion \citep{quintana02}.
\label{alignment}}
\end{figure}

We compared the orbital inclinations of the stellar companions ($i$) and of the transiting planets (assumed to be $90^\circ$, i.e. edge-on to our line of sight) to determine the planet-binary orbital alignment ($\sin|90-i|$) in each system. Note that this is only the minimum alignment, because we do not know the longitude of the ascending node of the transiting planet. Figure~\ref{alignment} shows a histogram of the planet-binary orbital alignment for our \nsample binary host systems. The uncertainties for each histogram bin were found by varying each binary inclination within it's Gaussian uncertainty over $10^5$ iterations, then taking the standard deviation of the values in each histogram bin from all iterations. In the case of asymmetric uncertainties in inclination, the larger uncertainty value was used. We found that our binary host orbits are more often aligned with the planetary orbits, with all mutual inclinations less than $60^\circ$.  Our result is consistent with the results of \citet{dupuy22} using linear orbital motion estimates and of \citet{christian22} using Gaia astrometric parameters. Specifically, \citet{christian22} found that only systems with $a < 700$~AU were preferentially aligned, so our sample confirms this result down to systems with much smaller separations.  

Furthermore, low mutual inclination between the planetary and binary orbits is consistent with theories of binary star formation and with planet formation in multi-star systems. Close binaries (such as those in this study), that formed in-situ via disk fragmentation or via turbulent fragmentation and migration, are expected to have binary orbits aligned with the primary stars’ protoplanetary disks. From a planet formation perspective, all of the binaries in our sample have mutual inclinations less than $60^\circ$, which is consistent with theoretical predictions. For example, \citep{quintana02} simulated planet formation around each star in the $\alpha$~Cen AB system; they found that planets could form more easily when the protoplanetary disk was inclined by $30-45^\circ$ compared to the binary orbit, but were unstable when the disk was inclined by $60^\circ$. Because most of our systems are well aligned, they likely did not undergo strong tidal interactions that would have torqued the protoplanetary disk and resulted in either non-transiting planets or poor binary-planet alignment.

\subsection{Planet Stability}
\begin{figure}
\centering
\includegraphics[width=0.48\textwidth]{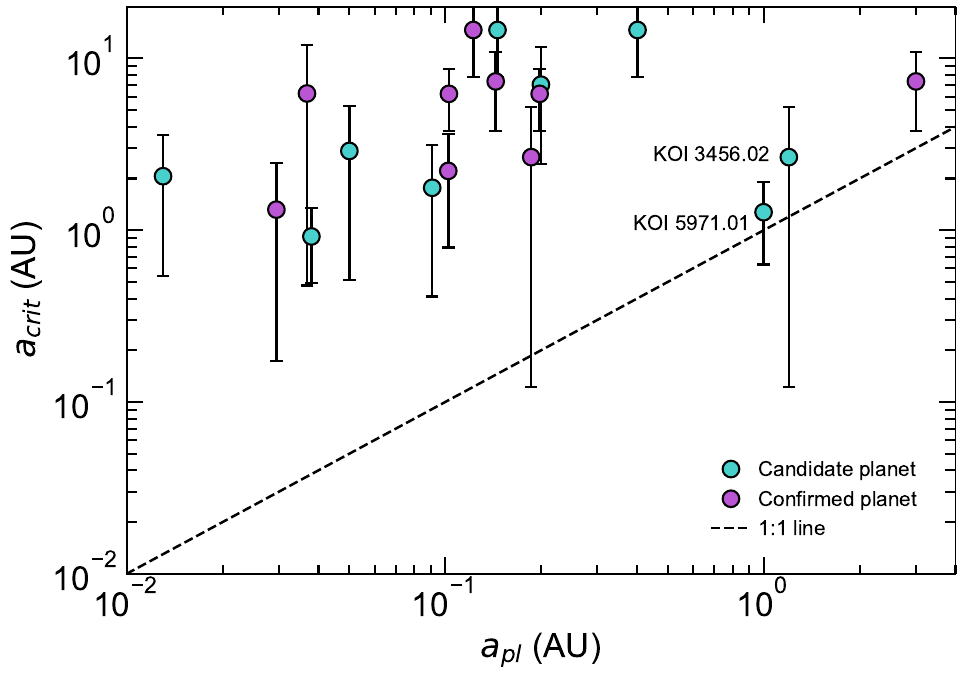}
\caption{Critical semi-major axis ($a_{crit}$) versus planet semi-major axis ($a_{pl}$) for the planet candidates in our sample. \citet{holman99} predict that planets with $a_{crit} > a_{pl}$ would be dynamically stable in binary systems.  All systems in our sample lie above the 1:1 line (dotted) and therefore are expected to be dynamically stable, though two planets (labeled) are close to their critical separation.
\label{acrit}}
\end{figure}

We next tested the binary and planet configurations of our sample against dynamical stability predictions from numerical simulations. We calculated the critical planet semi-major axis ($a_{crit}$) using Equation~1 in \citet{holman99}, for which planets with semi-major axes ($a_{pl}$) less than $a_{crit}$ would be stable orbiting one star in a binary system  over thousands of binary orbital cycles. For the multi-planet systems, we evaluated each planet separately. The critical value depends on the binary's semi-major axis, eccentricity, and mass ratio, so we used the mass ratios estimated from the speckle magnitude difference in Section~\ref{orbits}. The uncertainties in $a_{crit}$ were estimated by varying the binary parameters within their uncertainties over $10^5$ iterations and taking the standard deviation of the results. Figure~\ref{acrit} compares the planet separations to the critical separations for the binaries in our sample. We found that all planets have separations less than $a_{crit}$ and therefore would be dynamically stable. The only systems with planet separations near the critical separation are KOI~5971.01, with $a_{pl} \approx 1.0$~AU and $a_{crit} = 1.3\pm0.6$~AU, and KOI~3456.02, with $a_{pl} \approx 1.2$~AU and $a_{crit} = 2.7\pm2.1$~AU. These systems would benefit from continued speckle monitoring to better constrain the binary orbits and confirm $a_{crit}$, as well as additional transit follow-up to confirm the planetary nature of these planet candidates. Increasing the number of binary planet hosts in our sample and extending to longer period planets would provide additional tests of these dynamical stability models.

\section{Conclusions \label{conclusion}} 
We presented new relative astrometry of \nsample planet host binary systems and measured preliminary visual orbits using the \texttt{orbitize!} code. We investigated the mutual orbital inclination between the binary orbits and the transiting planets, and found that our binary host stars have orbital inclinations similar to those of the planets. Our result for close ($a<100$ AU) binaries is consistent with past work for wide, planet host binaries \citep[e.g.,][]{christian22}, and supports the predictions of planet formation simulations that binary companions inclined with respect to the protoplanetary disk will hinder planet formation.

We plan to continue monitoring our full sample of 40 planet host binaries in order to increase our orbital coverage and sample size and better constrain all of the orbital parameters. Eccentric companions cause increased torque on the protoplanetary disks and could cause the planets to become misaligned relative to the stellar companion, so investigating planet-binary orbital alignment as a function of binary separation and eccentricity would be a valuable test of planet formation theory. Continued astrometric monitoring will better constrain the binary orbital parameters (e.g., $i$ and $e$) and enable such investigations. We also started spectroscopic monitoring of these systems to measure the radial velocity trends and help break the degeneracy between the binary orbital inclination and eccentricity. Overall, we are working to build the orbital demographics of planet host binaries to better understand how planets form in multi-star systems.

Future work could also investigate the alignment of all components of the binary and planetary system, such as the spin-orbital alignment of the host star compared to the planetary and stellar companions. 
This would complement past work that typically studied planet-star alignment in single star systems \citep[e.g.,][]{winn10, triaud10, morton14} and star-companion alignment in non-planet hosting binaries \citep[e.g.,][]{albrecht07, justesen20}, as well as theoretical modeling of misaligned disks in binary systems \citep{lai14, martin14}. Such an investigation would require the Rossiter-McLaughlin technique to measure the orientation of the planet's orbit with respect to the host star's rotation \citep{albrecht22}, as well as measurement of the binary orbital inclination with respect to the stellar rotation. One could estimate the stellar rotation angle based on the rotation period, projected rotational velocity, and radius for the stars with rotational spot modulation in the light curve \citep{justesen20}. The spin-orbit alignment of planetary and binary systems is a useful probe of formation and dynamical history \citep{winn15}, so this work provides the binary orbital inclinations necessary for future studies.

\acknowledgments
The authors would like to thank the anonymous referee for their thorough review and helpful comments. We also thank the staff at Gemini and WIYN for their invaluable help conducting observations, as well as Josh Winn for useful conversations.

KVL is supported by an appointment to the NASA Postdoctoral Program at the NASA Ames Research Center, administered by Oak Ridge Associated Universities under contract with NASA. 
This work made use of the High-Resolution Imaging instruments NESSI, `Alopeke, and Zorro, which were funded by the NASA Exoplanet Exploration Program and built at the NASA Ames Research Center by Steve B. Howell, Nic Scott, Elliott P. Horch, and Emmett Quigley. 

Gemini Observatory is a program of NSF's NOIRLab, which is managed by the Association of Universities for Research in Astronomy (AURA) under a cooperative agreement with the National Science Foundation on behalf of the Gemini partnership: the National Science Foundation (United States), National Research Council (Canada), Agencia Nacional de Investigaci\'{o}n y Desarrollo (Chile), Ministerio de Ciencia, Tecnolog\'{i}a e Innovaci\'{o}n (Argentina), Minist\'{e}rio da Ci\^{e}ncia, Tecnologia, Inova\c{c}\~{o}es e Comunica\c{c}\~{o}es (Brazil), and Korea Astronomy and Space Science Institute (Republic of Korea). The authors wish to recognize and acknowledge the very significant cultural role and reverence that the summit of Maunakea has always had within the indigenous Hawaiian community. We are most fortunate to have the opportunity to conduct observations from this mountain. 

Data presented herein were obtained at the WIYN Observatory from telescope time allocated to NN-EXPLORE through the scientific partnership of the National Aeronautics and Space Administration, the National Science Foundation, and the NSF's National Optical-Infrared Astronomy Research Laboratory. The WIYN Observatory is a joint facility of the NSF's NOIRLab, Indiana University, the University of Wisconsin-Madison, Pennsylvania State University, the University of Missouri, the University of California-Irvine, and Purdue University. 

DRC and CAC acknowledge partial support from NASA Grant 18-2XRP18\_2-0007, and CAC acknowledges that this research was carried out at the Jet Propulsion Laboratory, California Institute of Technology, under a contract with the National Aeronautics and Space Administration (80NM0018D0004). 

This research has made use of the \citet{exofop} website and \citet{exoarchive}, which are operated by the California Institute of Technology, under contract with the National Aeronautics and Space Administration under the Exoplanet Exploration Program.

\facilities{Gemini North (`Alopeke), Gemini South (Zorro), WIYN (NESSI)}

\appendix

\begin{figure*}
\centering
\includegraphics[width=\textwidth]{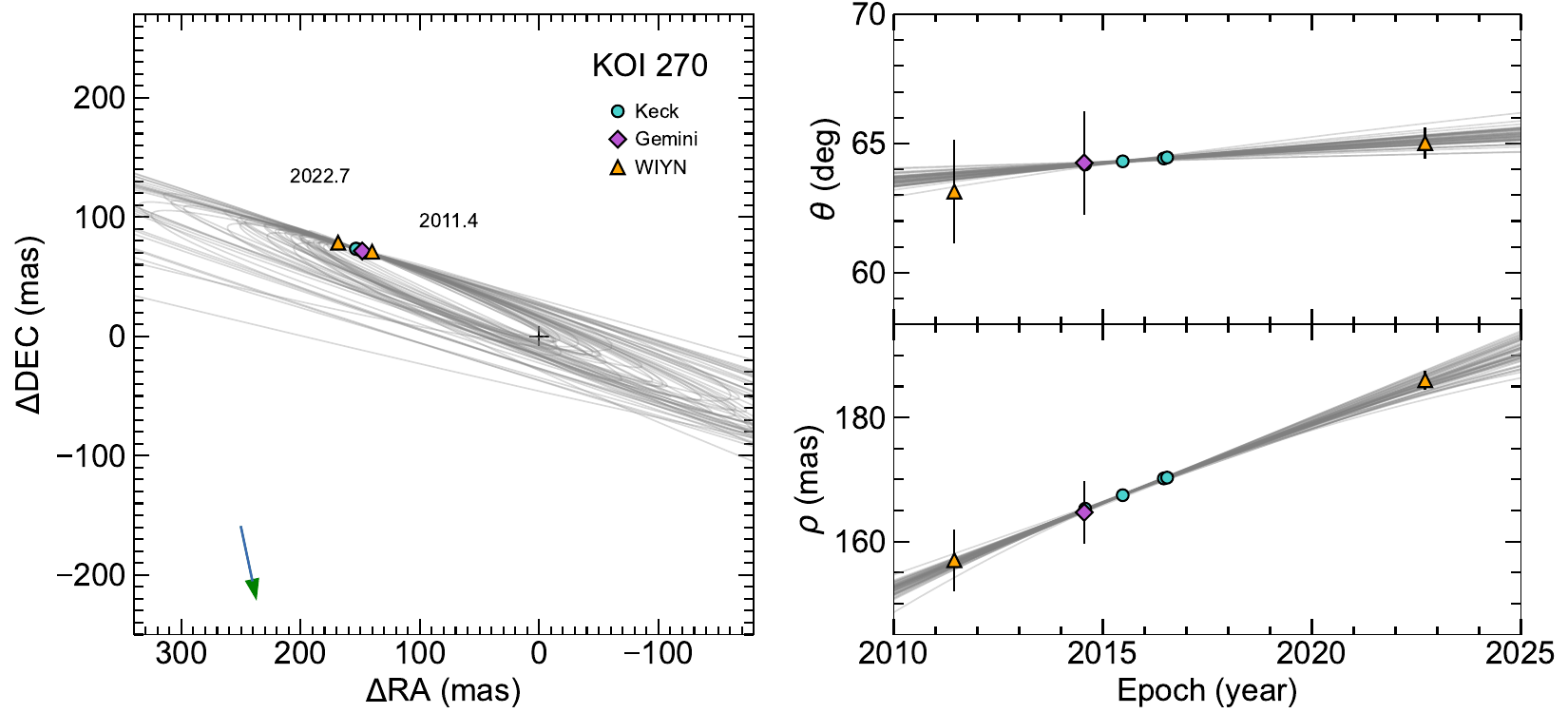}
\caption{\textit{Left:} Visual orbit solutions for KOI~270. The primary star is positioned at the origin (black cross), and the relative positions of the secondary component are marked with colored points. A random subset of the accepted orbital solutions from \texttt{orbitize!} are shown in grey. The green arrow shows the Gaia proper motion of the primary star (in a single year), which is inconsistent with the observed motion of the secondary. \textit{Right:} The observed changes in position angle and relative separation over time, plotted against the possible orbital solutions.
\label{vb_koi270}}
\end{figure*}

\begin{figure*}
\centering
\includegraphics[width=\textwidth]{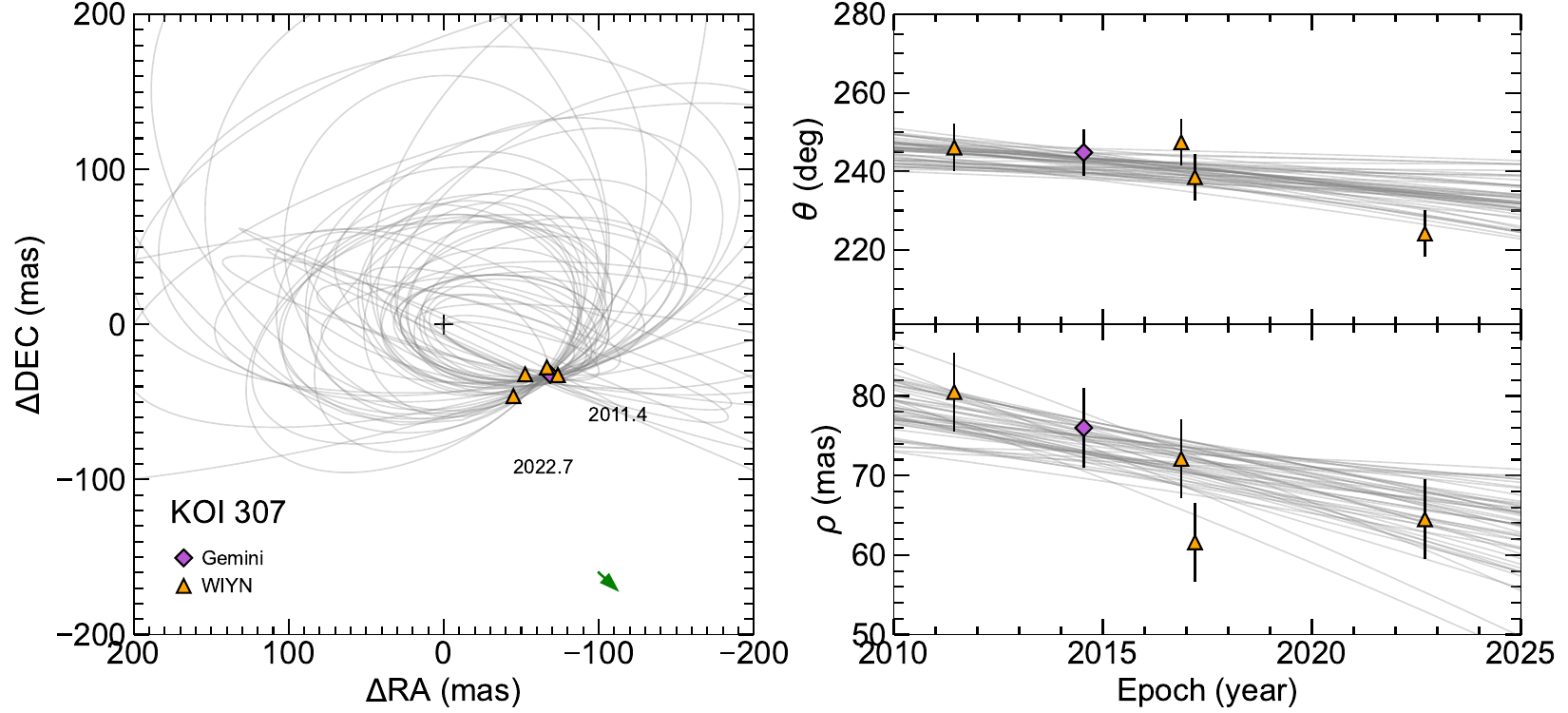}
\caption{\textit{Left:} Visual orbit solutions for KOI~307. The primary star is positioned at the origin (black cross), and the relative positions of the secondary component are marked with colored points. A random subset of the accepted orbital solutions from \texttt{orbitize!} are shown in grey.  The green arrow shows the Gaia proper motion of the primary star (in a single year), which is inconsistent with the observed motion of the secondary. \textit{Right:} The observed changes in position angle and relative separation over time, plotted against the possible orbital solutions.  
\label{vb_koi307}}
\end{figure*}

\begin{figure*}
\centering
\includegraphics[width=\textwidth]{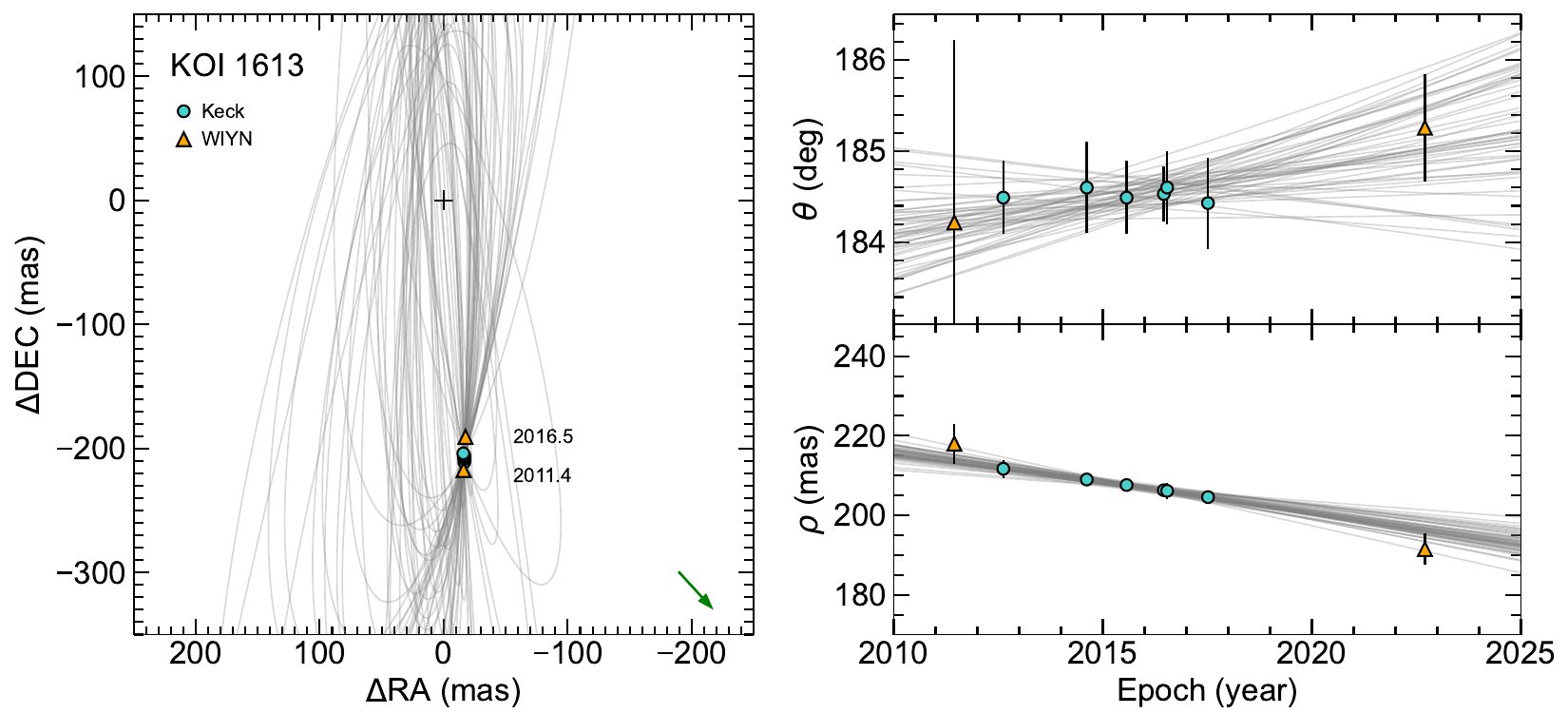}
\caption{\textit{Left:} Visual orbit solutions for KOI~1613. The primary star is positioned at the origin (black cross), and the relative positions of the secondary component are marked with colored points. A random subset of the accepted orbital solutions from \texttt{orbitize!} are shown in grey. The green arrow shows the Gaia proper motion of the primary star (in a single year), which is inconsistent with the observed motion of the secondary.  \textit{Right:} The observed changes in position angle and relative separation over time, plotted against the possible orbital solutions.
\label{vb_koi1613}}
\end{figure*}

\begin{figure*}
\centering
\includegraphics[width=\textwidth]{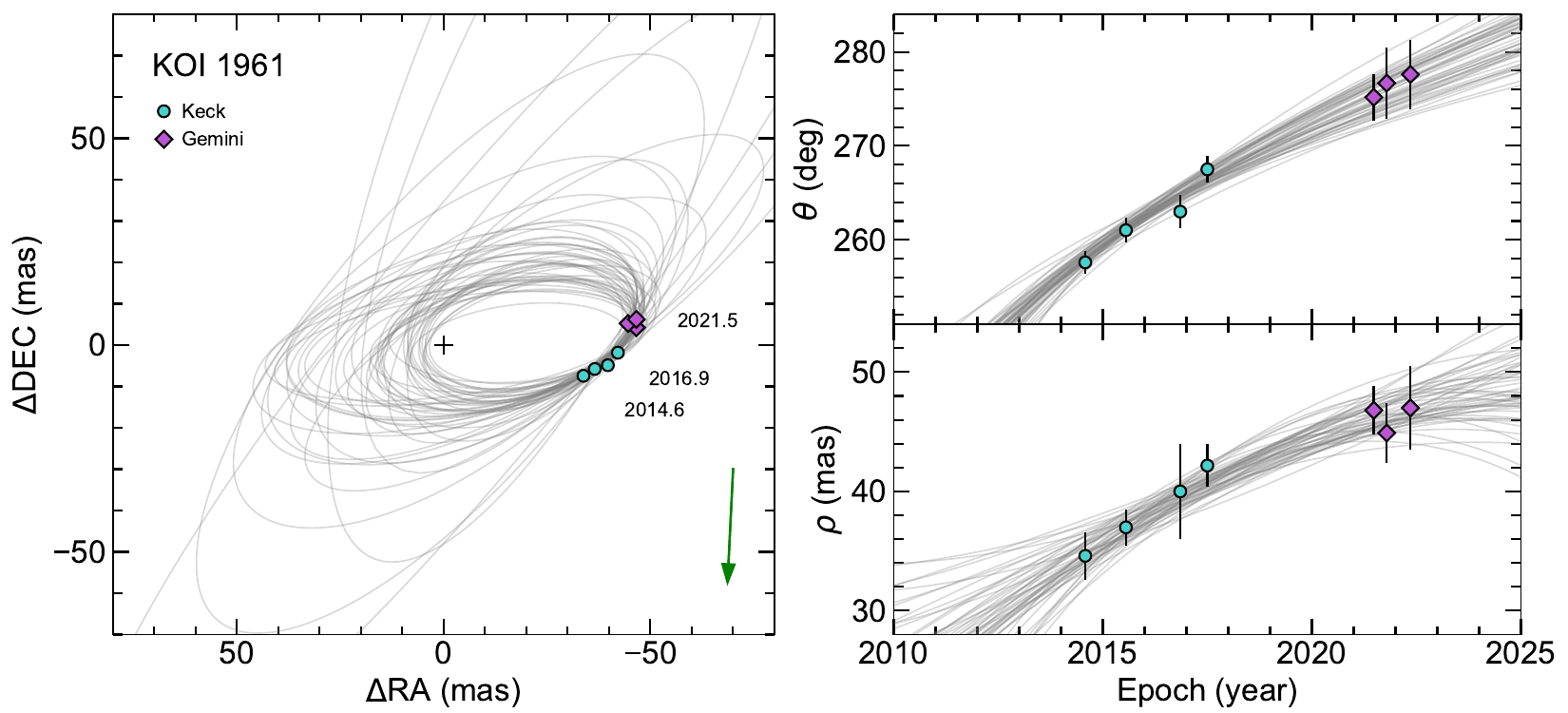}
\caption{\textit{Left:} Visual orbit solutions for KOI~1961. The primary star is positioned at the origin (black cross), and the relative positions of the secondary component are marked with colored points. A random subset of the accepted orbital solutions from \texttt{orbitize!} are shown in grey.  The green arrow shows the Gaia proper motion of the primary star (in a single year), which is inconsistent with the observed motion of the secondary. \textit{Right:} The observed changes in position angle and relative separation over time, plotted against the possible orbital solutions. 
\label{vb_koi1961}}
\end{figure*}

\begin{figure*}
\centering
\includegraphics[width=\textwidth]{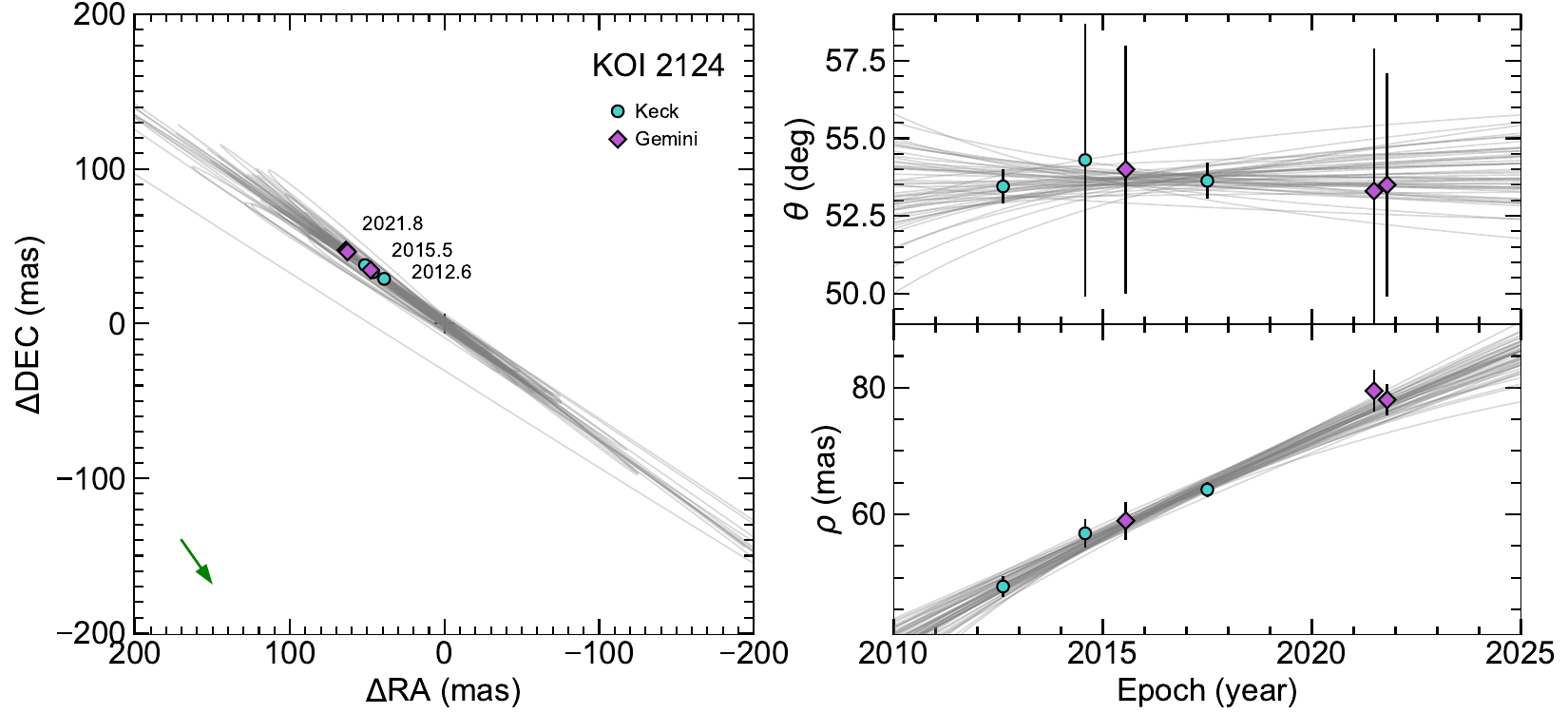}
\caption{\textit{Left:} Visual orbit solutions for KOI~2124. The primary star is positioned at the origin (black cross), and the relative positions of the secondary component are marked with colored points. A random subset of the accepted orbital solutions from \texttt{orbitize!} are shown in grey.  The green arrow shows the Gaia proper motion of the primary star (in a single year), which is inconsistent with the observed motion of the secondary. \textit{Right:} The observed changes in position angle and relative separation over time, plotted against the possible orbital solutions.
\label{vb_koi2124}}
\end{figure*}

\begin{figure*}
\centering
\includegraphics[width=\textwidth]{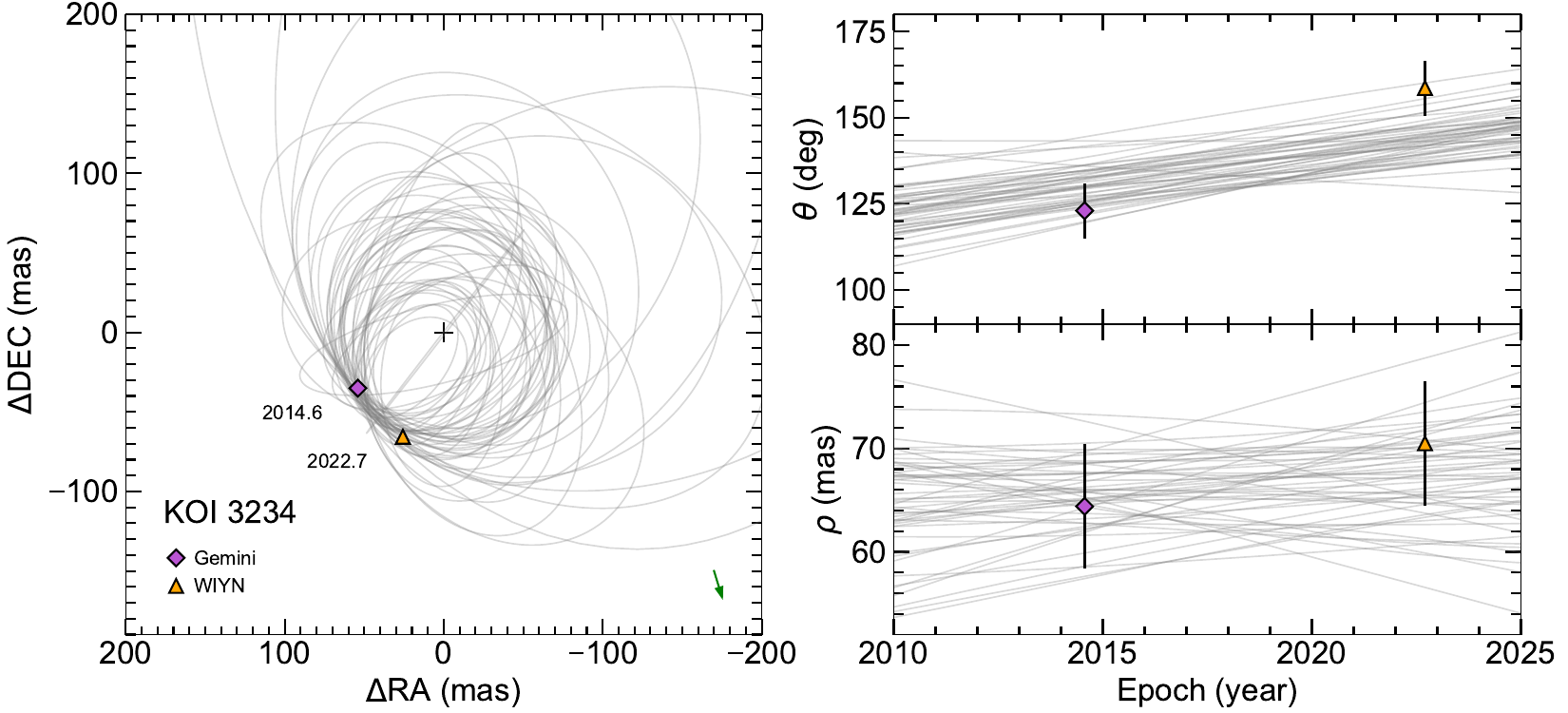}
\caption{\textit{Left:} Visual orbit solutions for KOI~3234. The primary star is positioned at the origin (black cross), and the relative positions of the secondary component are marked with colored points. A random subset of the accepted orbital solutions from \texttt{orbitize!} are shown in grey.  The green arrow shows the Gaia proper motion of the primary star (in a single year), which is inconsistent with the observed motion of the secondary. \textit{Right:} The observed changes in position angle and relative separation over time, plotted against the possible orbital solutions. 
\label{vb_koi3234}}
\end{figure*}

\begin{figure*}
\centering
\includegraphics[width=\textwidth]{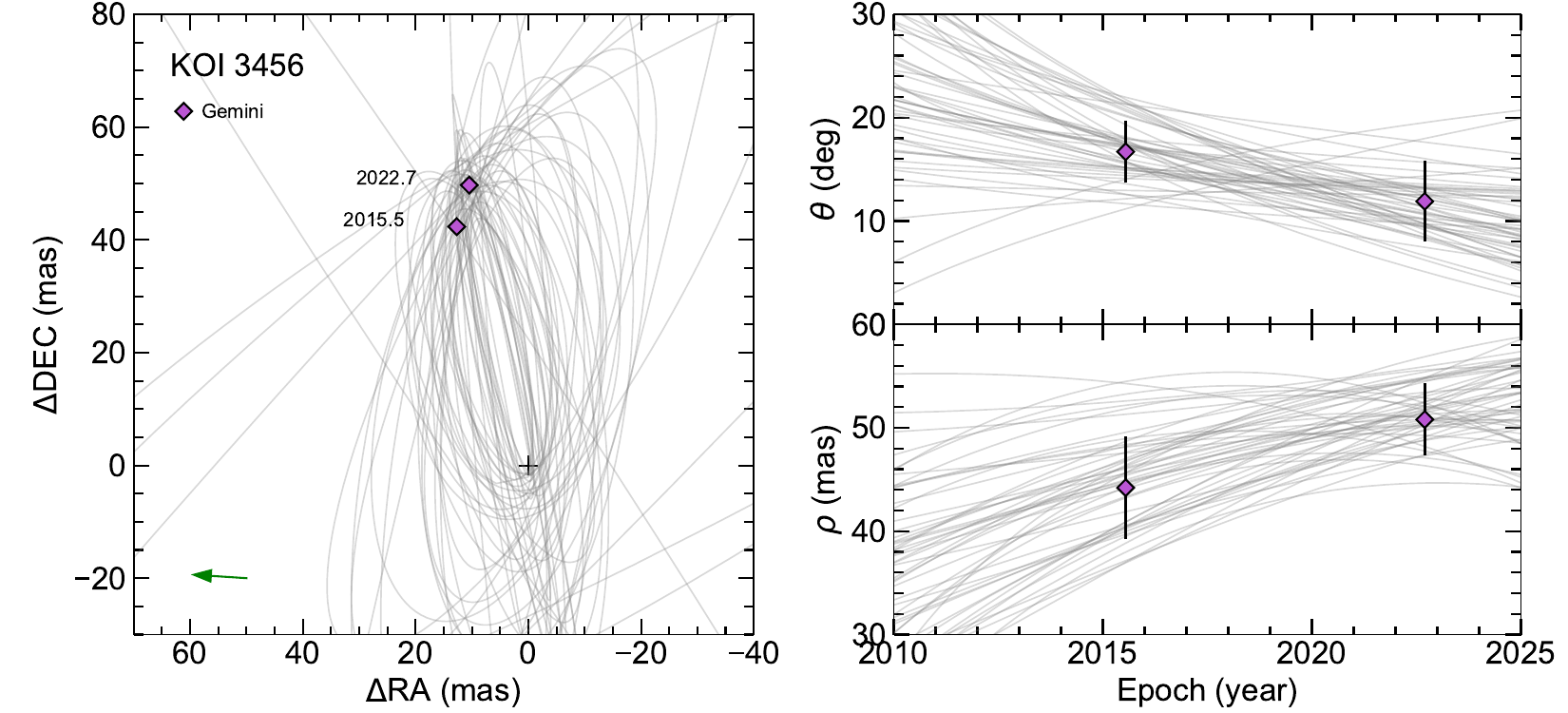}
\caption{\textit{Left:} Visual orbit solutions for KOI~3456. The primary star is positioned at the origin (black cross), and the relative positions of the secondary component are marked with colored points. A random subset of the accepted orbital solutions from \texttt{orbitize!} are shown in grey.  The green arrow shows the Gaia proper motion of the primary star (in a single year), which is inconsistent with the observed motion of the secondary. \textit{Right:} The observed changes in position angle and relative separation over time, plotted against the possible orbital solutions. 
\label{vb_koi3456}}
\end{figure*}

\begin{figure*}
\centering
\includegraphics[width=\textwidth]{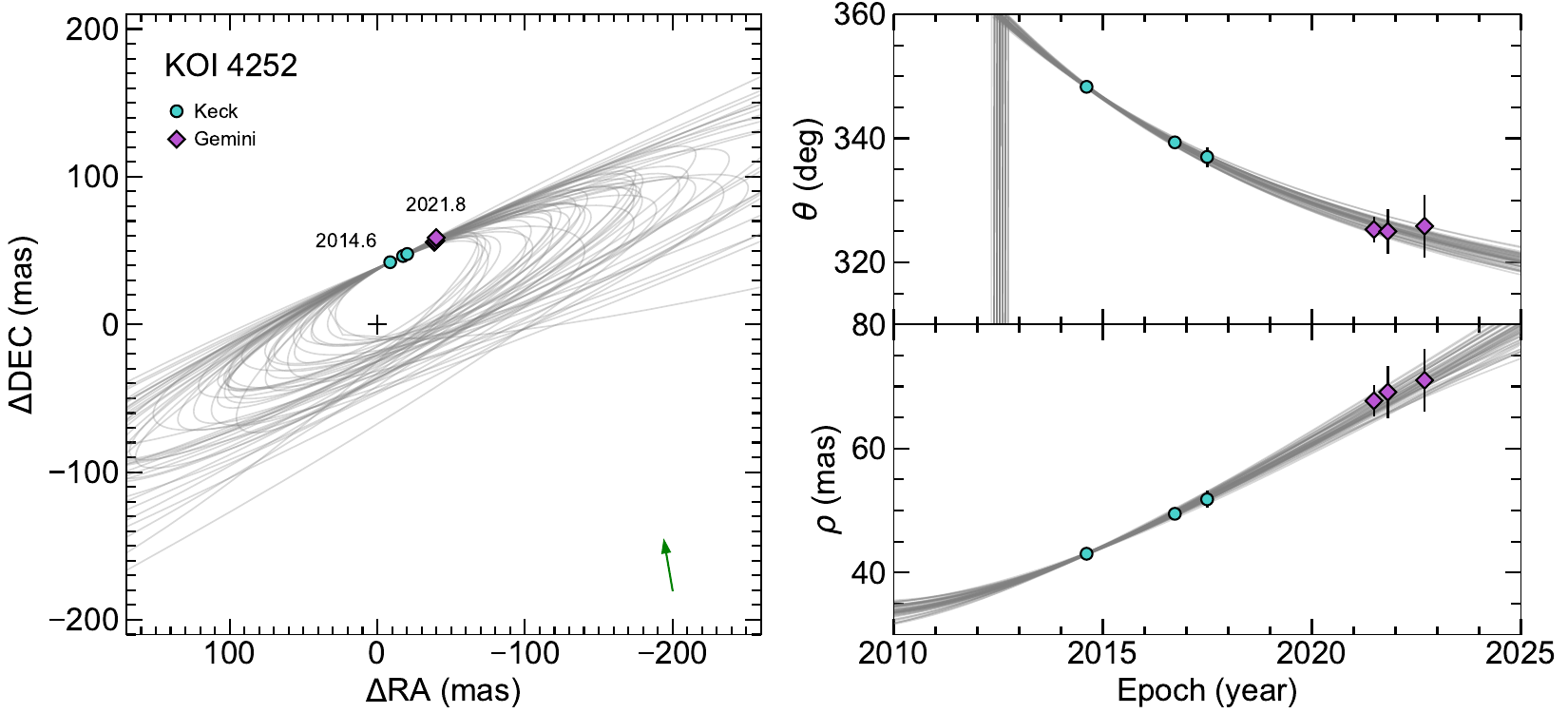}
\caption{\textit{Left:} Visual orbit solutions for KOI~4252. The primary star is positioned at the origin (black cross), and the relative positions of the secondary component are marked with colored points. A random subset of the accepted orbital solutions from \texttt{orbitize!} are shown in grey.  The green arrow shows the Gaia proper motion of the primary star (in a single year), which is inconsistent with the observed motion of the secondary. \textit{Right:} The observed changes in position angle and relative separation over time, plotted against the possible orbital solutions. 
\label{vb_koi4252}}
\end{figure*}

\begin{figure*}
\centering
\includegraphics[width=\textwidth]{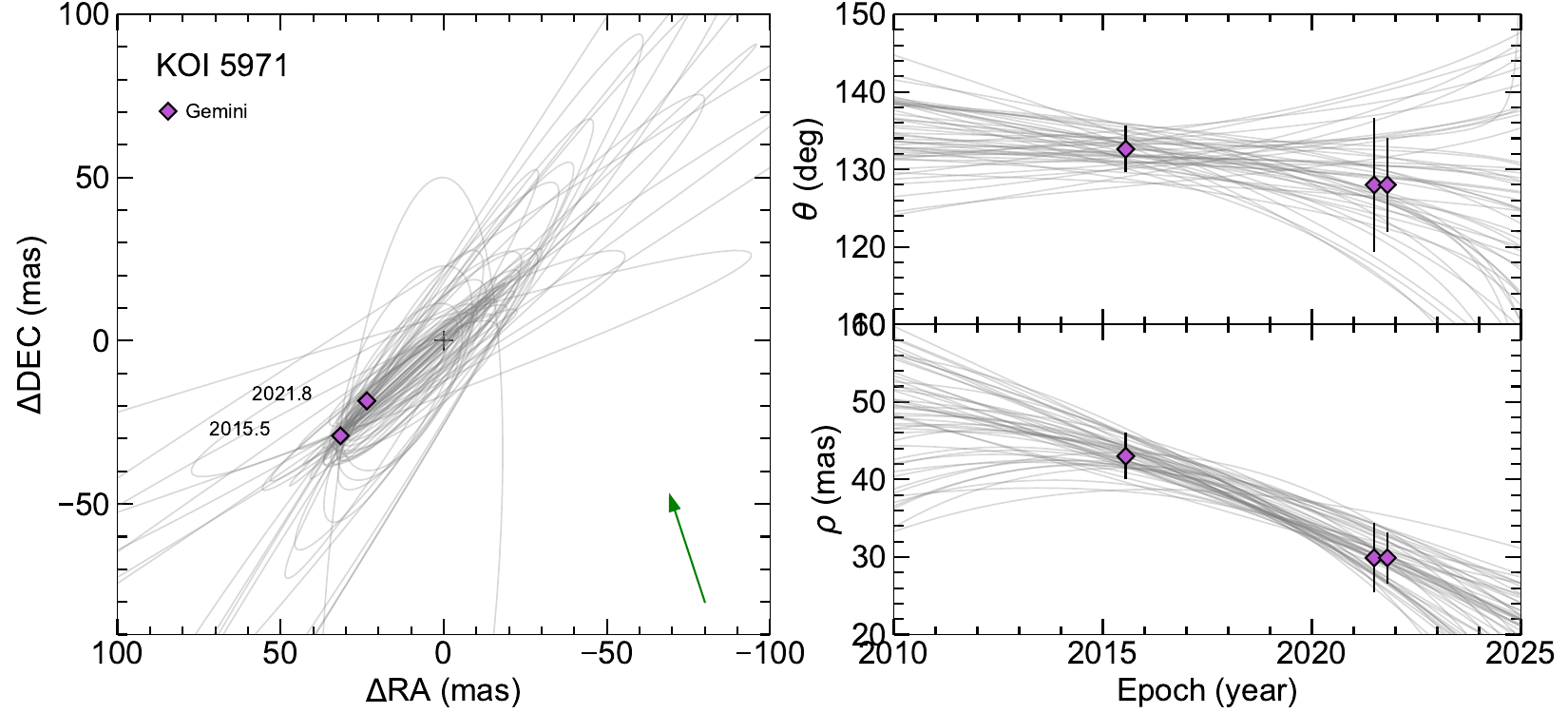}
\caption{\textit{Left:} Visual orbit solutions for KOI~5971. The primary star is positioned at the origin (black cross), and the relative positions of the secondary component are marked with colored points. A random subset of the accepted orbital solutions from \texttt{orbitize!} are shown in grey.  The green arrow shows the Gaia proper motion of the primary star (in a single year), which is inconsistent with the observed motion of the secondary. \textit{Right:} The observed changes in position angle and relative separation over time, plotted against the possible orbital solutions. 
\label{vb_koi5971}}
\end{figure*}

\begin{figure*}
\centering
\includegraphics[width=\textwidth]{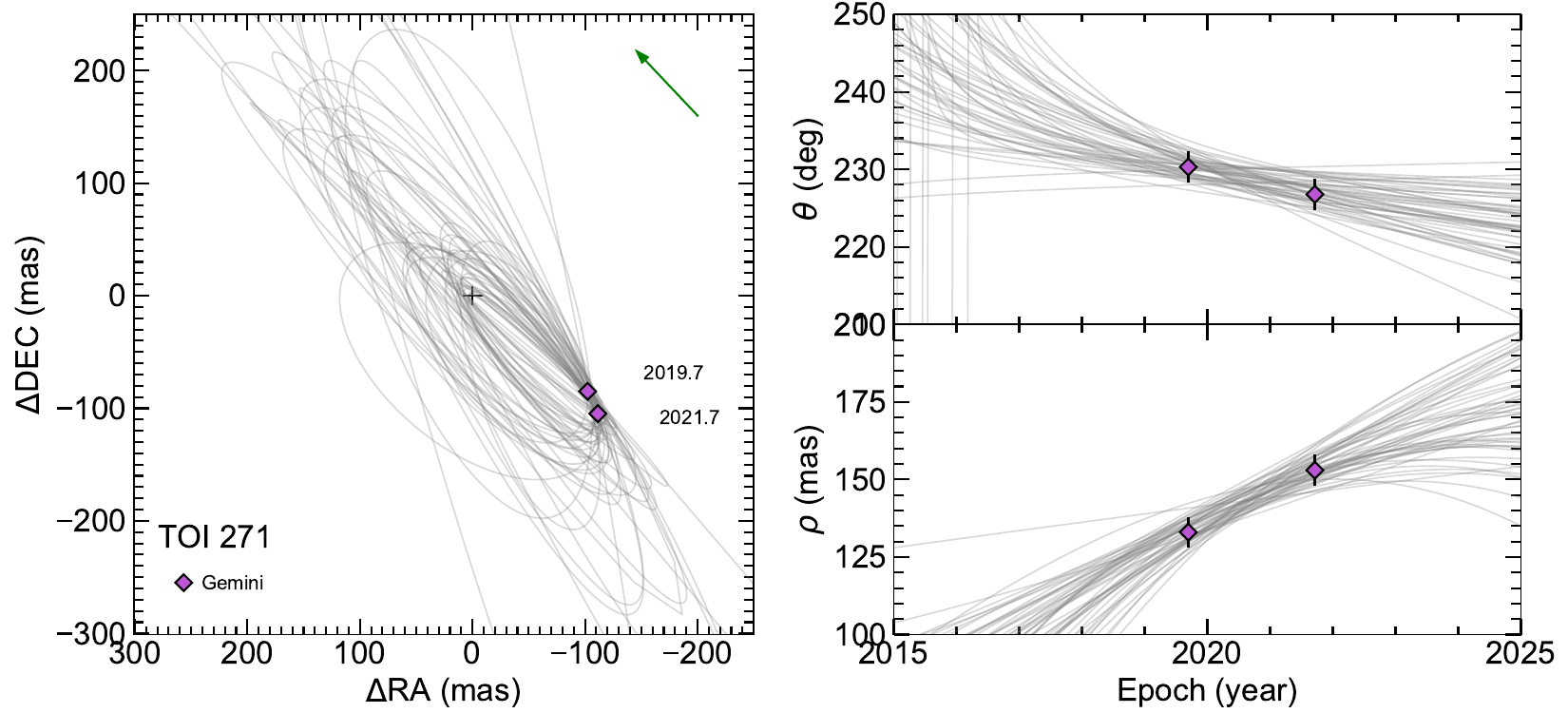}
\caption{\textit{Left:} Visual orbit solutions for TOI~271. The primary star is positioned at the origin (black cross), and the relative positions of the secondary component are marked with colored points. A random subset of the accepted orbital solutions from \texttt{orbitize!} are shown in grey.  The green arrow shows the Gaia proper motion of the primary star (in a single year), which is inconsistent with the observed motion of the secondary. \textit{Right:} The observed changes in position angle and relative separation over time, plotted against the possible orbital solutions. 
\label{vb_toi270}}
\end{figure*}

\begin{figure*}
\centering
\includegraphics[width=\textwidth]{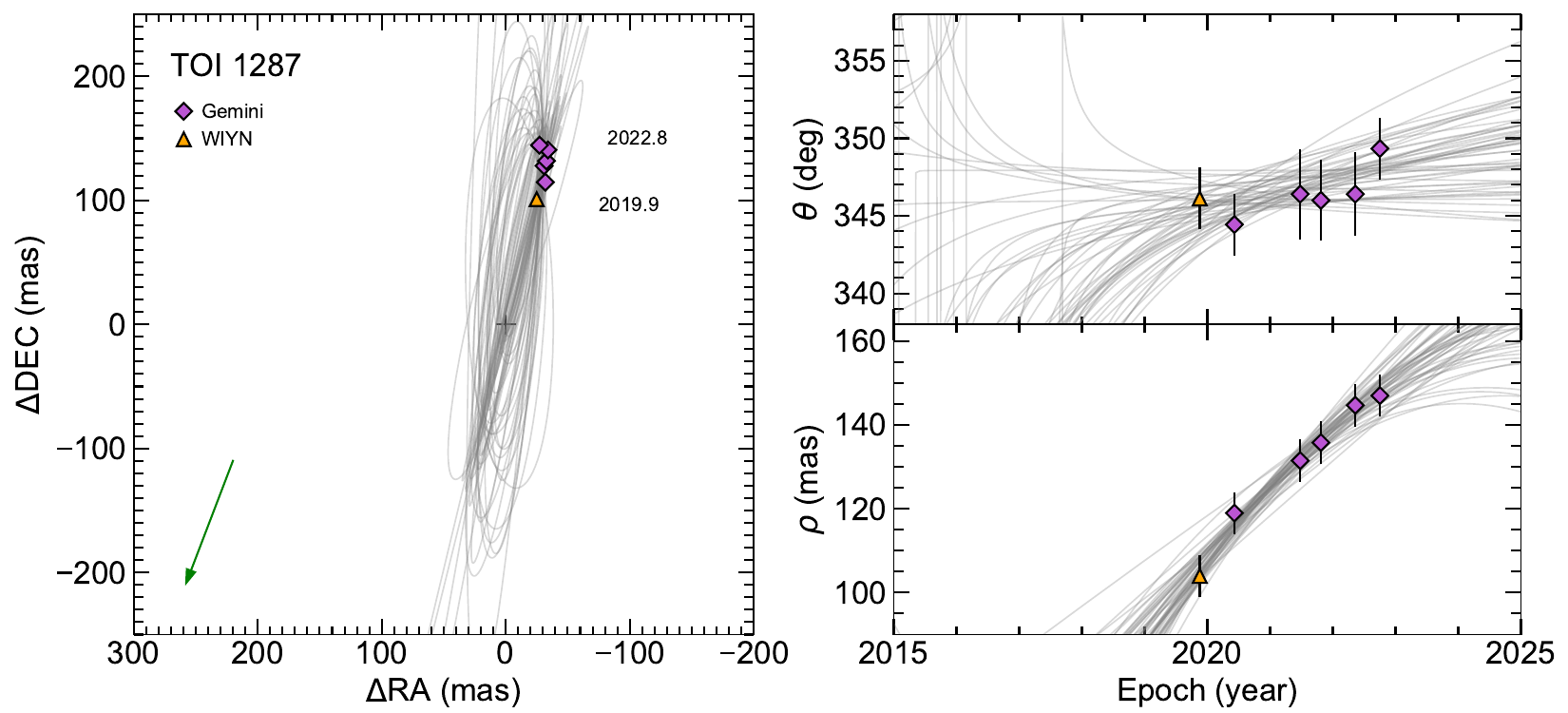}
\caption{\textit{Left:} Visual orbit solutions for TOI~1287. The primary star is positioned at the origin (black cross), and the relative positions of the secondary component are marked with colored points. A random subset of the accepted orbital solutions from \texttt{orbitize!} are shown in grey.  The green arrow shows the Gaia proper motion of the primary star (in a single year), which is inconsistent with the observed motion of the secondary. \textit{Right:} The observed changes in position angle and relative separation over time, plotted against the possible orbital solutions. 
\label{vb_toi1287}}
\end{figure*}

\begin{figure*}
\centering
\includegraphics[width=\textwidth]{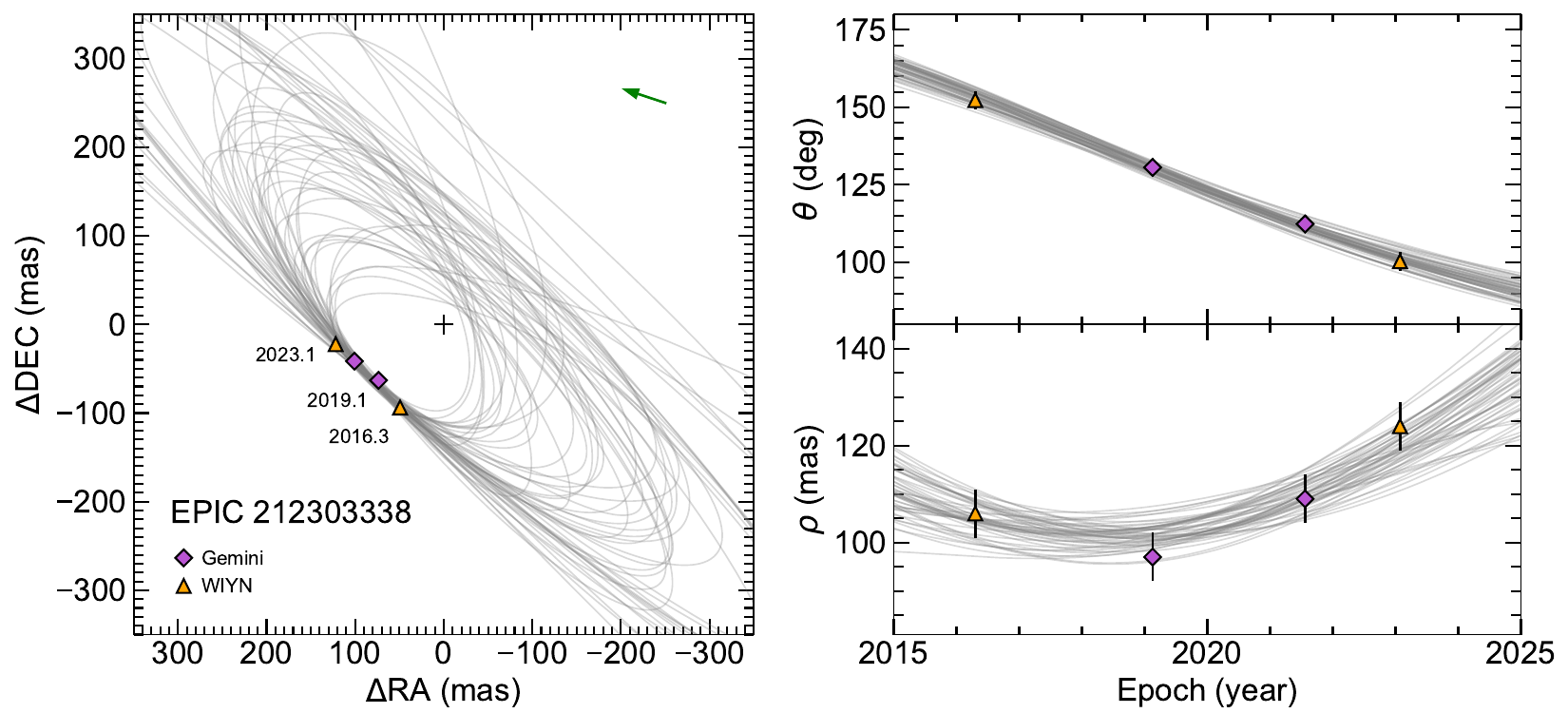}
\caption{\textit{Left:} Visual orbit solutions for EPIC~212303338. The primary star is positioned at the origin (black cross), and the relative positions of the secondary component are marked with colored points. A random subset of the accepted orbital solutions from \texttt{orbitize!} are shown in grey. The green arrow shows the Gaia proper motion of the primary star (in a single year), which is inconsistent with the observed motion of the secondary.  \textit{Right:} The observed changes in position angle and relative separation over time, plotted against the possible orbital solutions.
\label{vb_epic338}}
\end{figure*}

\begin{figure*}
\centering
\includegraphics[width=\textwidth]{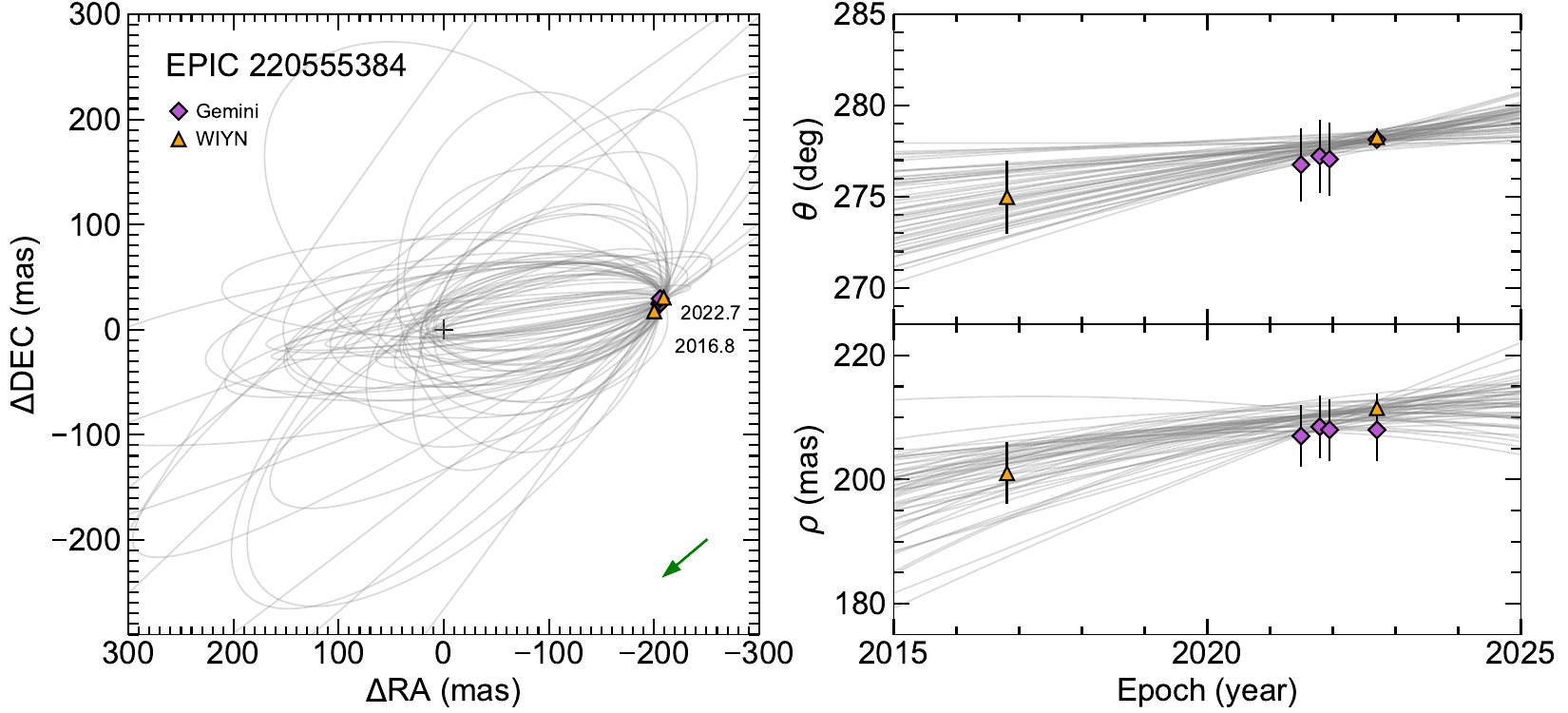}
\caption{\textit{Left:} Visual orbit solutions for EPIC~220555384. The primary star is positioned at the origin (black cross), and the relative positions of the secondary component are marked with colored points. A random subset of the accepted orbital solutions from \texttt{orbitize!} are shown in grey. The green arrow shows the Gaia proper motion of the primary star (in a single year), which is inconsistent with the observed motion of the secondary. \textit{Right:} The observed changes in position angle and relative separation over time, plotted against the possible orbital solutions.
\label{vb_epic384}}
\end{figure*}

\break



\begin{thebibliography}{}

\bibitem[Albrecht et al.(2007)]{albrecht07} Albrecht, S., Reffert, S., Snellen, I., et al.\ 2007, \aap, 474, 565. doi:10.1051/0004-6361:20077953

\bibitem[Albrecht et al.(2022)]{albrecht22} Albrecht, S.~H., Dawson, R.~I., \& Winn, J.~N.\ 2022, \pasp, 134, 082001. doi:10.1088/1538-3873/ac6c09

\bibitem[Artymowicz \& Lubow(1994)]{artymowicz94} Artymowicz, P. \& Lubow, S.~H.\ 1994, \apj, 421, 651. doi:10.1086/173679

\bibitem[Batalha et al.(2013)]{batalha13} Batalha, N.~M., Rowe, J.~F., Bryson, S.~T., et al.\ 2013, \apjs, 204, 24. doi:10.1088/0067-0049/204/2/24

\bibitem[Behmard et al.(2022)]{behmard22} Behmard, A., Dai, F., \& Howard, A.~W.\ 2022, \aj, 163, 160. doi:10.3847/1538-3881/ac53a7

\bibitem[Bergfors et al.(2013)]{bergfors13} Bergfors, C., Brandner, W., Daemgen, S., et al.\ 2013, \mnras, 428, 182. doi:10.1093/mnras/sts019

\bibitem[Blunt et al.(2017)]{ofti} Blunt, S., Nielsen, E.~L., De Rosa, R.~J., et al.\ 2017, \aj, 153, 229. doi:10.3847/1538-3881/aa6930

\bibitem[Blunt et al.(2020)]{orbitize} Blunt, S., Wang, J.~J., Angelo, I., et al.\ 2020, \aj, 159, 89. doi:10.3847/1538-3881/ab6663

\bibitem[Cadman et al.(2022)]{cadman22} Cadman, J., Hall, C., Fontanive, C., et al.\ 2022, \mnras, 511, 457. doi:10.1093/mnras/stac033

\bibitem[Christian et al.(2022)]{christian22} Christian, S., Vanderburg, A., Becker, J., et al.\ 2022, \aj, 163, 207. doi:10.3847/1538-3881/ac517f

\bibitem[Chontos et al.(2022)]{chontos22} Chontos, A., Murphy, J.~M.~A., MacDougall, M.~G., et al.\ 2022, \aj, 163, 297. doi:10.3847/1538-3881/ac6266

\bibitem[Clark et al.(2022)]{clark22} Clark, C.~A., van Belle, G.~T., Ciardi, D.~R., et al.\ 2022, \aj, 163, 232. doi:10.3847/1538-3881/ac6101

\bibitem[Colton et al.(2021)]{colton21} Colton, N.~M., Horch, E.~P., Everett, M.~E., et al.\ 2021, \aj, 161, 21. doi:10.3847/1538-3881/abc9af

\bibitem[Dawson \& Johnson(2018)]{dawson18} Dawson, R.~I. \& Johnson, J.~A.\ 2018, \araa, 56, 175. doi:10.1146/annurev-astro-081817-051853

\bibitem[Dupuy et al.(2016)]{dupuy16} Dupuy, T.~J., Kratter, K.~M., Kraus, A.~L., et al.\ 2016, \apj, 817, 80. doi:10.3847/0004-637X/817/1/80

\bibitem[Dupuy et al.(2022)]{dupuy22} Dupuy, T.~J., Kraus, A.~L., Kratter, K.~M., et al.\ 2022, \mnras, 512, 648. doi:10.1093/mnras/stac306

\bibitem[Everett et al.(2015)]{everett15} Everett, M.~E., Barclay, T., Ciardi, D.~R., et al.\ 2015, \aj, 149, 55. doi:10.1088/0004-6256/149/2/55

\bibitem[Exoplanet Follow-up Observing Program(2022)]{exofop} Exoplanet Follow-up Observing Program. 2022, IPAC. doi:10.26134/EXOFOP5

\bibitem[Fontanive et al.(2019)]{fontanive19} Fontanive, C., Rice, K., Bonavita, M., et al.\ 2019, \mnras, 485, 4967. doi:10.1093/mnras/stz671

\bibitem[Fontanive \& Bardalez Gagliuffi(2021)]{fontanive21} Fontanive, C. \& Bardalez Gagliuffi, D.\ 2021, Frontiers in Astronomy and Space Sciences, 8, 16. doi:10.3389/fspas.2021.625250

\bibitem[Furlan et al.(2017)]{furlan17} Furlan, E., Ciardi, D.~R., Everett, M.~E., et al.\ 2017, \aj, 153, 71. doi:10.3847/1538-3881/153/2/71

\bibitem[Gaia Collaboration et al.(2016)]{gaia} Gaia Collaboration, Prusti, T., de Bruijne, J.~H.~J., et al.\ 2016, \aap, 595, A1. doi:10.1051/0004-6361/201629272

\bibitem[Gaia Collaboration et al.(2022)]{gaiaDR3} Gaia Collaboration, Vallenari, A., Brown, A.~G.~A., et al.\ 2022, arXiv:2208.00211. doi:10.48550/arXiv.2208.00211

\bibitem[Guerrero et al.(2021)]{toi} Guerrero, N.~M., Seager, S., Huang, C.~X., et al.\ 2021, \apjs, 254, 39. doi:10.3847/1538-4365/abefe1

\bibitem[Harris et al.(2012)]{harris12} Harris, R.~J., Andrews, S.~M., Wilner, D.~J., et al.\ 2012, \apj, 751, 115. doi:10.1088/0004-637X/751/2/115

\bibitem[Hatzes et al.(2003)]{hatzes03} Hatzes, A.~P., Cochran, W.~D., Endl, M., et al.\ 2003, \apj, 599, 1383. doi:10.1086/379281

\bibitem[Hirsch et al.(2017)]{hirsch17} Hirsch, L.~A., Ciardi, D.~R., Howard, A.~W., et al.\ 2017, \aj, 153, 117. doi:10.3847/1538-3881/153/3/117

\bibitem[Hirsch et al.(2021)]{hirsch21} Hirsch, L.~A., Rosenthal, L., Fulton, B.~J., et al.\ 2021, \aj, 161, 134. doi:10.3847/1538-3881/abd639

\bibitem[Holman \& Wiegert(1999)]{holman99} Holman, M.~J. \& Wiegert, P.~A.\ 1999, \aj, 117, 621. doi:10.1086/300695

\bibitem[Horch et al.(2011)]{horch11} Horch, E.~P., Gomez, S.~C., Sherry, W.~H., et al.\ 2011, \aj, 141, 45. doi:10.1088/0004-6256/141/2/45 

\bibitem[Horch et al.(2014)]{horch14} Horch, E.~P., Howell, S.~B., Everett, M.~E., et al.\ 2014, \apj, 795, 60. doi:10.1088/0004-637X/795/1/60

\bibitem[Howell et al.(2011)]{howell11} Howell, S.~B., Everett, M.~E., Sherry, W., et al.\ 2011, AJ, 142, 19. doi:10.1088/0004-6256/142/1/19  

\bibitem[Howell et al.(2021)]{howell21} Howell, S.~B., Matson, R.~A., Ciardi, D.~R., et al.\ 2021, \aj, 161, 164. doi:10.3847/1538-3881/abdec6

\bibitem[Huber et al.(2016)]{epic} Huber, D., Bryson, S.~T., Haas, M.~R., et al.\ 2016, \apjs, 224, 2. doi:10.3847/0067-0049/224/1/2

\bibitem[Jang-Condell(2015)]{jc15} Jang-Condell, H.\ 2015, \apj, 799, 147. doi:10.1088/0004-637X/799/2/147

\bibitem[Justesen \& Albrecht(2020)]{justesen20} Justesen, A.~B. \& Albrecht, S.\ 2020, \aap, 642, A212. doi:10.1051/0004-6361/202039138

\bibitem[Kostov et al.(2019)]{kostov19} Kostov, V.~B., Mullally, S.~E., Quintana, E.~V., et al.\ 2019, \aj, 157, 124. doi:10.3847/1538-3881/ab0110

\bibitem[Kraus et al.(2012)]{kraus12} Kraus, A.~L., Ireland, M.~J., Hillenbrand, L.~A., et al.\ 2012, \apj, 745, 19. doi:10.1088/0004-637X/745/1/19

\bibitem[Kraus et al.(2016)]{kraus16} Kraus, A.~L., Ireland, M.~J., Huber, D., et al.\ 2016, \aj, 152, 8. doi:10.3847/0004-6256/152/1/8

\bibitem[Kruse et al.(2019)]{kruse19} Kruse, E., Agol, E., Luger, R., et al.\ 2019, \apjs, 244, 11. doi:10.3847/1538-4365/ab346b

\bibitem[Lai (2014)]{lai14} Lai, D.\ 2014, \mnras, 440, 3532. doi:10.1093/mnras/stu485

\bibitem[Lester et al.(2021)]{lester21} Lester, K.~V., Matson, R.~A., Howell, S.~B., et al.\ 2021, \aj, 162, 75. doi:10.3847/1538-3881/ac0d06

\bibitem[Martin et al.(2014)]{martin14} Martin, R.~G., Nixon, C., Lubow, S.~H., et al.\ 2014, \apjl, 792, L33. doi:10.1088/2041-8205/792/2/L33

\bibitem[Matson et al.(2018)]{matson18} Matson, R.~A., Howell, S.~B., Horch, E.~P., et al.\ 2018, AJ, 156, 31. doi:10.3847/1538-3881/aac778

\bibitem[Moe \& Kratter(2021)]{moe21} Moe, M. \& Kratter, K.~M.\ 2021, \mnras, 507, 3593. doi:10.1093/mnras/stab2328

\bibitem[Morton \& Winn(2014)]{morton14} Morton, T.~D. \& Winn, J.~N.\ 2014, \apj, 796, 47. doi:10.1088/0004-637X/796/1/47

\bibitem[Morton et al.(2016)]{morton16} Morton, T.~D., Bryson, S.~T., Coughlin, J.~L., et al.\ 2016, \apj, 822, 86. doi:10.3847/0004-637X/822/2/86

\bibitem[Mullally et al.(2015)]{q1-16table} Mullally, F., Coughlin, J.~L., Thompson, S.~E., et al.\ 2015, \apjs, 217, 31. doi:10.1088/0067-0049/217/2/31

\bibitem[Muterspaugh et al.(2006)]{muterspaugh06} Muterspaugh, M.~W., Lane, B.~F., Konacki, M., et al.\ 2006, \aap, 446, 723. doi:10.1051/0004-6361:20053749

\bibitem[Muterspaugh et al.(2010)]{muterspaugh10} Muterspaugh, M.~W., Hartkopf, W.~I., Lane, B.~F., et al.\ 2010, \aj, 140, 1623. doi:10.1088/0004-6256/140/6/1623

\bibitem[NASA Exoplanet Archive(2019)]{exoarchive} NASA Exoplanet Archive. 2019, IPAC, Confirmed Planets Table. doi:10.26133/NEA1

\bibitem[Offner et al.(2010)]{offner10} Offner, S.~S.~R., Kratter, K.~M., Matzner, C.~D., et al.\ 2010, \apj, 725, 1485. doi:10.1088/0004-637X/725/2/1485

\bibitem[Pecaut \& Mamajek(2013)]{pecaut13} Pecaut, M.~J. \& Mamajek, E.~E.\ 2013, \apjs, 208, 9

\bibitem[Quintana et al.(2002)]{quintana02} Quintana, E.~V., Lissauer, J.~J., Chambers, J.~E., et al.\ 2002, \apj, 576, 982. doi:10.1086/341808

\bibitem[Raghavan et al.(2010)]{raghavan10} Raghavan, D., McAlister, H.~A., Henry, T.~J., et al.\ 2010, \apjs, 190, 1. doi:10.1088/0067-0049/190/1/1

\bibitem[Rowe et al.(2015)]{q1-12table} Rowe, J.~F., Coughlin, J.~L., Antoci, V., et al.\ 2015, \apjs, 217, 16. doi:10.1088/0067-0049/217/1/16

\bibitem[Schaefer et al.(2016)]{schaefer16} Schaefer, G.~H., Hummel, C.~A., Gies, D.~R., et al.\ 2016, \aj, 152, 213. doi:10.3847/0004-6256/152/6/213 

\bibitem[Scott et al.(2018)]{scott18} Scott, N.~J., Howell, S.~B., Horch, E.~P., et al.\ 2018, PASP, 130, 054502. doi:10.1088/1538-3873/aab484

\bibitem[Scott et al.(2021)]{scott21} Scott, N.~J., Howell, S.~B., Gnilka, C.~L., et al.\ 2021, Frontiers in Astronomy and Space Sciences, 8, 138. doi:10.3389/fspas.2021.716560

\bibitem[Su et al.(2021)]{su21} Su, X.-N., Xie, J.-W., Zhou, J.-L., et al.\ 2021, \aj, 162, 272. doi:10.3847/1538-3881/ac2ba3

\bibitem[Thebault \& Haghighipour(2015)]{thebault15} Thebault, P. \& Haghighipour, N.\ 2015, Planetary Exploration and Science: Recent Results and Advances, 309. doi:10.1007/978-3-662-45052-9\_13

\bibitem[Triaud et al.(2010)]{triaud10} Triaud, A.~H.~M.~J., Collier Cameron, A., Queloz, D., et al.\ 2010, \aap, 524, A25. doi:10.1051/0004-6361/201014525

\bibitem[Tokovinin et al.(2015)]{tokovinin15} Tokovinin, A., Mason, B.~D., Hartkopf, W.~I., et al.\ 2015, \aj, 150, 50. doi:10.1088/0004-6256/150/2/50

\bibitem[Valizadegan et al.(2022)]{valizadegan22} Valizadegan, H., Martinho, M.~J.~S., Wilkens, L.~S., et al.\ 2022, \apj, 926, 120. doi:10.3847/1538-4357/ac4399

\bibitem[Van Eylen \& Albrecht(2015)]{ve15} Van Eylen, V. \& Albrecht, S.\ 2015, \apj, 808, 126. doi:10.1088/0004-637X/808/2/126

\bibitem[Wang et al.(2014a)]{wang14a} Wang, J., Xie, J.-W., Barclay, T., et al.\ 2014, \apj, 783, 4. doi:10.1088/0004-637X/783/1/4  

\bibitem[Wang et al.(2014b)]{wang14b} Wang, J., Fischer, D.~A., Xie, J.-W., et al.\ 2014, \apj, 791, 111. doi:10.1088/0004-637X/791/2/111  

\bibitem[Winn et al.(2010)]{winn10} Winn, J.~N., Fabrycky, D., Albrecht, S., et al.\ 2010, \apjl, 718, L145. doi:10.1088/2041-8205/718/2/L145

\bibitem[Winn \& Fabrycky(2015)]{winn15} Winn, J.~N. \& Fabrycky, D.~C.\ 2015, \araa, 53, 409. doi:10.1146/annurev-astro-082214-122246

\bibitem[Winters et al.(2019a)]{winters19} Winters, J.~G., Henry, T.~J., Jao, W.-C., et al.\ 2019, \aj, 157, 216. doi:10.3847/1538-3881/ab05dc

\bibitem[Winters et al.(2019b)]{winters19b} Winters, J.~G., Medina, A.~A., Irwin, J.~M., et al.\ 2019, \aj, 158, 152. doi:10.3847/1538-3881/ab364d

\bibitem[Ziegler et al.(2020)]{ziegler20} Ziegler, C., Tokovinin, A., Brice{\~n}o, C., et al.\ 2020, \aj, 159, 19. doi:10.3847/1538-3881/ab55e9

\bibitem[Zurlo et al.(2021)]{zurlo21} Zurlo, A., Cieza, L.~A., Ansdell, M., et al.\ 2021, \mnras, 501, 2305. doi:10.1093/mnras/staa3674


\end{thebibliography}
\end{document}